\documentclass[10pt,aps,prd,floats,superscriptaddress,amsfonts,notitlepage,nofootinbib,eqsecnum,twocolumn]{revtex4-1}\pdfoutput=1
\usepackage{ifpdf}
\usepackage[utf8]{inputenc} % bibtex related
\usepackage[UKenglish]{babel}
\usepackage[final]{graphicx}
\usepackage{hyperref}
\usepackage{amsmath, amsthm, amssymb}
\usepackage{url}
\usepackage{color}
\usepackage{array}
\usepackage[caption=false]{subfig}
\usepackage{dcolumn}
	\newcolumntype{.}{D{.}{.}{13}}
	\newcolumntype{d}[1]{D{.}{.}{#1}}
\usepackage{longtable}
\allowdisplaybreaks[1]
\newcommand{\ord}{\mathcal{O}}
\newcommand{\la}{\langle}
\newcommand{\ra}{\rangle}

\newcommand{\be}{\begin{equation}}
\newcommand{\ee}{\end{equation}}
\newcommand{\ba}{\begin{eqnarray}}
\newcommand{\ea}{\end{eqnarray}}
\newcommand{\bi}{\begin{itemize}}
\newcommand{\ei}{\end{itemize}}
\newcommand{\bef}{\begin{frame}}
\newcommand{\ef}{\end{frame}}
\newcommand{\bs}{\begin{small}}
\newcommand{\es}{\end{small}}
\newcommand{\Lim}[1]{\raisebox{0.5ex}{\scalebox{0.8}{$\displaystyle \lim_{#1}\;$}}}
\newcommand\T{\rule{0pt}{2.6ex}}       % Top strut for vertical spacing in tables
\newcommand\B{\rule[-1.2ex]{0pt}{0pt}} % Bottom strut
\newcommand{\AT}{ALT}
\newcommand{\BDG}{BDG}

\newcommand{\abs}[1]{\lvert#1\rvert}			%Absolute value
\newcommand{\avg}[1]{\langle #1\rangle }		%Average

\newcommand{\hh}[1]{\left(#1\right) }			%Scaling parentheses
\newcommand{\bh}[1]{\bigl(#1\bigr) }			%big parentheses
\newcommand{\Bh}[1]{\Bigr(#1\Bigr) }			%Big parentheses
\newcommand{\id}[1]{\operatorname{d}\!#1}													%Differential
\newcommand{\npd}[3]{\frac{\operatorname{\partial^#3}\!#1}{\operatorname{\partial}\!#2^#3}}	%nth Partial derivative

\newcommand{\nsd}{(\text{---})}		%no significant digits

\usepackage{titlesec}
\titleclass{\subsubsubsection}{straight}[\subsection]

\newcounter{subsubsubsection}[subsubsection]

\renewcommand\thesubsubsubsection{\thesubsubsection.\arabic{subsubsubsection}}

\titleformat{\subsubsubsection}
  {\normalfont\normalsize}{\thesubsubsubsection}{1em}{} % {\normalfont\normalsize\bfseries}{\thesubsubsubsection}{1em}{} %original version
\titlespacing*{\subsubsubsection}
{0pt}{3.25ex plus 1ex minus .2ex}{1.5ex plus .2ex}

\makeatletter
\renewcommand\paragraph{\@startsection{paragraph}{5}{\z@}%
  {3.25ex \@plus1ex \@minus.2ex}%
  {-1em}%
  {\normalfont\normalsize\bfseries}}
\renewcommand\subparagraph{\@startsection{subparagraph}{6}{\parindent}
  {3.25ex \@plus1ex \@minus .2ex}%
  {-1em}%
  {\normalfont\normalsize\bfseries}}
\def\toclevel@subsubsubsection{4}
\def\toclevel@paragraph{5}
\def\toclevel@paragraph{6}
\def\l@subsubsubsection{\@dottedtocline{4}{7em}{4em}}
\def\l@paragraph{\@dottedtocline{5}{10em}{5em}}
\def\l@subparagraph{\@dottedtocline{6}{14em}{6em}}
\makeatother

\makeatletter
\@addtoreset{subsubsubsection}{section}
\@addtoreset{subsubsubsection}{subsection}
\makeatother

\begin{document}

\title{Numerical computation of the effective-one-body potential \texorpdfstring{$q$}{} using self-force results}

\author{Sarp Akcay}
\email{Sarp.Akcay@ucd.ie}
\affiliation{School of Mathematical Sciences and Complex \& Adaptive Systems Laboratory, University College Dublin, Belfield, Dublin 4, Ireland}
\author{Maarten van de Meent}
\email{M.vandeMeent@soton.ac.uk}
\affiliation{Mathematical Sciences, University of Southampton, Southampton, SO17 1BJ, United Kingdom}
\affiliation{STAG Research Centre, University of Southampton, Southampton, SO17 1BJ, United Kingdom}

%\date{\today}

\begin{abstract}
The effective-one-body theory (EOB) describes the conservative dynamics of compact binary systems in terms of an effective Hamiltonian approach. The Hamiltonian for moderately eccentric motion of two non-spinning compact objects in the extreme mass-ratio limit 
is given in terms of three potentials: $a(v), \bar{d}(v), q(v)$. By generalizing the first law of mechanics for (non-spinning) black hole binaries to eccentric orbits, [A. Le Tiec \prd{\bf92}, 084021 (2015)] recently obtained new expressions for $\bar{d}(v)$ and $q(v)$ in terms of quantities that can be readily computed using
the gravitational self-force approach.
Using these expressions we present a new computation of the EOB potential $q(v)$ by combining results from two independent numerical self-force codes.
We determine $q(v)$ for inverse binary separations in the range $1/1200 \le v \lesssim 1/6$. Our computation thus provides the first-ever strong-field results for $q(v)$.
We also obtain $\bar{d}(v)$ in our entire domain to a fractional accuracy of $\gtrsim 10^{-8}$. We find that our results are compatible with the known post-Newtonian expansions for $\bar{d}(v)$ and $q(v)$ in the weak field, and agree with previous (less accurate) numerical results for $\bar{d}(v)$ in the strong field.
\end{abstract}	

%\pacs{}
%\keywords{}

\maketitle

\section{Introduction}\label{sec:intro}
%{\bf Attempt at a more compact introduction; anybody interested in this fairly technical result is going to be bored sick of the usual blabla}
The last few years have seen an increasing synergy between the various approaches used to solve the two-body problem in general relativity, extending the relevance of each approach well beyond its usual domain of validity. For example, %data gathered using gravitational self-force (GSF) methods expanding the equations of motion in the ratio of the masses of the components of the binary 
input from the gravitational self-force (GSF) appoach, which is based on an expansion of the equations of motion in the mass ratio of a compact binary system,
has been instrumental in fixing an ambiguous parameter in the recent derivation of fourth order post-Newtonian (pN) equations of motion \cite{JaSc.13,BiDa.13,Da.al.14,Jaranowski:2015lha}. The effective-one-body (EOB) formalism \cite{BuDa.99,BuDa.00} 
sits at the center of this synergestic activity drawing input from self-force, post-Newtonian, and numerical relativity calculations to provide a computationally effective method for calculating gravitational wave templates for compact binary mergers \cite{Pan:2011gk,Taracchini:2012ig,Taracchini:2013rva,Bernuzzi:2014owa,Nagar:2015xqa}. 

The aim of this paper is to utilize recent technological advances in eccentric-orbit self-force computations \cite{Akcay:2015pza,Maarten2015} to determine the linear-in-mass-ratio contributions to the potentials in the EOB Hamiltonian for moderately eccentric, 
non-spinning binaries. Currently, the numerical relativity calibrated EOB-based wave templates (EOBNR \cite{Pan:2011gk,Taracchini:2012ig,Taracchini:2013rva}) --- in use in the detection pipeline of the Advanced LIGO detector --- only model quasicircular inspirals. 
Since eccentric (comparable mass) binaries are of considerable interest as gravitational wave sources \cite{We.03,OL.al.09,Th.11,AnPe.12,KoLe.12,Sa.al.14}, improving the accuracy of EOB models for eccentric binaries is essential.
The main focus of this work will be to determine these potentials in the strong-field regime, where their pN expansions (currently known up to fourth order \cite{DJS2015}) are insufficient to reliably describe the two-body dynamics. %expected to fail.

The key idea is to use a relation between the EOB potentials and the so-called ``redshift (pseudo) invariant''\footnote{The redshift is a ``pseudo-invariant'' rather than a true gauge invariant, since it is invariant only under a restricted class of gauge transformations \cite{De.08,Sa.al.08,BaSa.11}.} recently found by using the first law of mechanics for compact binaries on eccentric orbits \cite{LeTiec2015}. The GSF correction to the redshift for eccentric orbits was first calculated by Barack and Sago \cite{Barack:2011ed}. Much-improved results have recently been produced by the authors with frequency-domain methods using both a Lorenz-gauge approach \cite{Akcay:2015pza}, and a radiation-gauge approach where the metric perturbation is reconstructed from the Weyl scalars \cite{Maarten2015}. 
Combining these methods, we determine the EOB $\bar{d}$ and $q$ potentials for dimensionless binary separations of $6 \lesssim r \le 1200$ (see Sec.~\ref{sec:EOB_intro} for the precise definition of $r$). We then compare our results in the weak field with pN expressions from Refs.~\cite{DJS2015,BDA}. 
We also check our strong-field values for the $\bar{d}$ potential with published results of Ref.~\cite{Akcay_et_al2012} and unpublished results of Ref.~\cite{Nori_pvt}.

This paper is organized as follows. In Sec.~\ref{sec:EOB} we review EOB in the extreme mass-ratio regime and display the relations between $\bar{d}(v)$ and $q(v)$ and the redshift. In Sec.~\ref{sec:computation} we present the details of our numerical calculation. 
Sec.~\ref{sec:results} summarizes our results. Finally, in App.~\ref{sec:App_A} we present our entire numerical data for $\bar{d}(v)$ and~$q(v)$. Throughout this article we use $(-,+,+,+)$ for the metric signature and geometrized units $G=c=1$. 
Henceforth, we refer to Refs.~\cite{LeTiec2015} and \cite{BDA} as \AT\ (A. Le Tiec), and \BDG~(Bini-Damour-Geralico), respectively. Unless specified otherwise, all mentions of accuracy will imply \emph{relative} accuracy.

\section{Preliminaries}\label{sec:EOB}
\subsection{EOB formalism}\label{sec:EOB_intro}
We consider a bounded binary system consisting of two compact masses $m_1$ and $m_2$ moving in a mutually eccentric orbit. In the EOB formalism the conservative dynamics of this system is described by a Hamiltonian \cite{Buonanno:1998gg},
\begin{equation}\label{eq:EOBham}
H_\mathrm{EOB} = m\sqrt{1+2\nu\hh{\hat{H}_\mathrm{eff}-1}},
\end{equation}
where $m=m_1+m_2$ is the total mass of the system, $\nu = \frac{m_1 m_2}{m^2}$ is the symmetric mass ratio, and $\hat{H}_\mathrm{eff}$ is an effective Hamiltonian describing an effective particle with mass $\mu=m\nu$ moving in an effective spacetime with metric,
\begin{equation}\label{eq:effmetric}
 g_{\alpha\beta}^\mathrm{eff} = -A(r;\nu) \id{t}^2 + B(r;\nu)\id{r}^2 + r^2\id{\Omega}^2,
\end{equation}
where $r$ is the orbital separation of the binary in Schwarzschild-like coordinates. The effective Hamiltonian is given by \cite{Da.al.00,DJS2015}
\begin{equation}\label{eq:effham}
\begin{split}
&\hat{H}_\mathrm{eff}(r,p_r,L_0)=\\
&\quad\sqrt{A(r;\nu)\hh{\mu^2+\frac{L_0^2}{r^2}+\frac{p_r^2}{B(r;\nu)}+Q(r,p_r;\nu)}},
\end{split}
\end{equation}
where $L_0$ is the conserved angular momentum and $p_r$ is the canonical momentum conjugate to the effective particle's radial position $r$.

The effective metric \eqref{eq:effmetric} can be regarded as a deformed Schwarzschild metric with the symmetric mass ratio $\nu$ acting as a deformation parameter. In the limit that $\nu\to0$ and $p_r/\mu \to 0$, the EOB potentials $A$, $B$, and $Q$ can be written as
\begin{align}
	A(u;\nu) &=	1-2u+\nu a(u) + \ord(\nu^2)\\
	\bar{D}(u;\nu) &= \frac{1}{AB} = 	1+\nu \bar{d}(u) + \ord(\nu^2)\\
	Q(u;\nu) &= \mu^2\nu q(u) \Bh{\frac{p_r}{\mu}}^4 + \ord\left(\nu^2,\Bh{\frac{p_r}{\mu}}^6\right),
\end{align}
where we introduced the notation $u\equiv m/r$. The linear-in-mass-ratio potentials $a(u)$, $\bar{d}(u)$, and $q(u)$ can be studied in the small mass-ratio regime using GSF techniques. Data from self-force calculations [on circular orbits in Schwarzschild spacetime] of Refs.~\cite{Le.al2.12,Akcay_et_al2012,Ba.al.12}
have enabled the determination of $a(u)$ in the entire domain $0<u<1/3$. Using a relation between $a(u)$, $\bar{d}(u)$ and the self-force correction to the periapsis advance for slightly eccentric orbits \cite{Da.10}, 
Refs.~\cite{Akcay_et_al2012,Ba.al.12} were able to compute $\bar{d}(u)$ in the range $0<u<1/6$. Meanwhile, the potential $q(u)$ has only been determined in the weak-field regime up to 4pN \cite{DJS2015}. The main goal of this paper is to provide a strong-field 
computation of $q(u)$ using its relation with the redshift established by the first law of binary mechanics for eccentric orbits presented in \AT, which we review presently.

\subsection{EOB potentials from redshift}\label{sec:EOB_potentials}
The redshift (pseudo)invariant, first introduced by Detweiler \cite{Detweiler:2008ft} for circular orbits and later generalized to eccentric orbits by Barack and Sago \cite{Barack:2011ed}, is defined as
\begin{equation}\label{eq:defredshift}
U(\Omega_r,\Omega_\phi;\nu) \equiv \frac{T_r}{\mathcal{T}_r},
\end{equation}
where $\Omega_r$ and $\Omega_\phi$ are the radial and azimuthal frequencies of the orbit as measured in the locally regular conservative ``effective'' spacetime (see \cite{Barack:2011ed}), including all self-force corrections. Similarly, $T_r$ and $\mathcal{T}_r$ are 
the radial period measured in Boyer-Lindquist coordinate time and proper time respectively, again including all conservative self-force corrections. 
To define self-force correction to the redshift we expand Eq.~\eqref{eq:defredshift} in powers of the mass ratio\footnote{Traditionally in the self-force literature, expansions are done with respect to the mass ratio  $m_1/m_2$. 
Here, for the sake of convenient comparison with EOB literature, we write all expansions with respect to the \emph{symmetric} mass ratio $\nu$. Obviously, for $m_1 \ll m_2$ the two are equivalent at the leading order.}
\begin{equation}\label{eq:Unuexp}
U(\Omega_r,\Omega_\phi;\nu) = U_{(0)}(\Omega_r,\Omega_\phi) + \nu U_{(1)}(\Omega_r,\Omega_\phi)+ \ord(\nu^2),
\end{equation}
where the expansion is understood to happen at fixed frequencies $\Omega_r$ and $\Omega_\phi$. However, for the sake of computational convenience, we parametrize our orbits (and all quantities depending on them) using the inverse semi-latus rectum $v$ and eccentricity $e$, 
which in turn are defined from the periapsis $r_p$ and apapsis $r_a$ by
\begin{align}
	v	&= m\frac{r_a +r_p}{2\, r_a r_p},\quad\text{and}\\
	e	&= \frac{r_a - r_p}{r_a +r_p}.
\end{align}
We can relate $v,e$ to $u$ via Darwin's standard parametrization of bound motion $u(\chi) = v\,(1 + e\cos\chi)$ where $\chi \in[0,2\pi]$ is the relativistic anomaly \cite{Da.61}.

In Ref. \cite{Akcay:2015pza} it was shown that for small mass-ratio systems $U_{(1)}$ can be calculated from
\begin{equation}
U_{(1)}(v,e) = \frac{U_{(0)}(v,e)}{2} \avg{h_{uu}^{R}(v,e)}, 
\end{equation}
where $h_{uu}^{R}$ is the (Detweiler-Whiting) regularized metric perturbation \cite{Detweiler:2002mi} double-contracted with object 1's four-velocity $u^\mu_{(0)}$ which is defined with respect to the background spacetime generated by $m_2$, and $\avg{\cdot}$ indicates an orbital 
average with respect to proper time. In the small-$e$ limit the $\mathcal{O}(\nu)$ correction to the redshift $U$ can be written as an expansion in even powers of the eccentricity
\begin{equation}\label{eq:U_e2_expansion}
U_{(1)}(v,e) = U_{(1)}^{e^0}(v) +\frac{e^2}{2!}U_{(1)}^{e^2}(v)
 +\frac{e^4}{4!}U_{(1)}^{e^4}(v)+ \mathcal{O}(e^6),
\end{equation}
where
\begin{align}
U_{(1)}^{e^0}(v) &\equiv \Lim{e\to0}U_{(1)}(v,e),
\\
U_{(1)}^{e^2}(v)&\equiv \Lim{e\to0}\npd{U_{(1)}(v,e)}{e}{2},\quad\text{and}
\\
U_{(1)}^{e^4}(v)&\equiv\Lim{e\to0}\npd{U_{(1)}(v,e)}{e}{4}.
\end{align}
Analogous notation will be used for the $e^2$ expansions of other quantities.

In the following we are often interested in the inverse redshift,
\begin{equation}\label{eq:invredshift}
z(\Omega_r,\Omega_\phi;\nu) \equiv \frac{1}{U(\Omega_r,\Omega_\phi;\nu)},
\end{equation}
which has an expansion in the mass-ratio analogous to Eq.~\eqref{eq:Unuexp}. In particular,
\begin{align} \label{eq:z1ecc}
 z_{(1)}(v,e) &= -\frac{ U_{(1)}(v,e) }{U_{(0)}(v,e)^2}.
\end{align}
Using Eq.~\eqref{eq:z1ecc}, and the $e^2$ coefficients of $U_{(0)}(v,e)$, one straightforwardly obtains the small-$e$ expansion of $z_{(1)}(v)$ analogous to Eq.\eqref{eq:U_e2_expansion} with the coefficients given by
\begin{align}
&z_{(1)}^{e^0}(v) = -\frac{U_{(1)}^{e^0}}{(U_{(0)}^{e^0})^2} \label{eq:z1circ} ,\\
&z_{(1)}^{e^2}(v) = -\frac{U_{(1)}^{e^2}}{(U_{(0)}^{e^0})^2} + 2 \frac{U_{(1)}^{e^0}}{(U_{(0)}^{e^0})^3} U_{(0)}^{e^2}, \label{eq:z1e2}\\
&\begin{aligned}
z_{(1)}^{e^4}(v) = 
&\frac{12 U_{(1)}^{e^2}}{(U_{(0)}^{e^0})^3} 
U_{(0)}^{e^2}-\frac{U_{(1)}^{e^4}}{(U_{(0)}^{e^0})^2} \\
&\quad+ 2\frac{U_{(1)}^{e^0}}{(U_{(0)}^{e^0})^3}  \hh{ U_{(0)}^{e^4}-9\frac{ \bh{U_{(0)}^{e^2}}^2}{U_{(0)}^{e^0}}
},
\end{aligned}
\label{eq:z1e4}
\end{align}
where
\begin{align}
&U_{(0)}^{e^0} = \frac{1}{\sqrt{1-3v}},\\
 &U_{(0)}^{e^2}= -\frac{3 v \left(1-10v + 22v^2\right)}{(1-6v) (1-3v)^{3/2} (1-2v)}, \label{eq:Lim_e0_d2Ude2}\\
 &\begin{aligned}
 U_{(0)}^{e^4} = &-\frac{9 v^2 }{(1-6v)^3 (1-3v)^{5/2} (1-2v)^3}
 \\
  &\quad\times\bh{1-6 v-163 v^2+2188 v^3
  \\
  &\qquad-10565 v^4+22860 v^5-18612v^6}. \label{eq:Lim_e0_d4Ude4}
\end{aligned}
\end{align} 

Using its newly formulated first law for compact binaries on eccentric orbits, \AT~ derived expressions for $a(v)$, $\bar{d}$, and $q(v)$ in terms of $z_{(1)}^{e^0}(v)$, $z_{(1)}^{e^2}(v)$, $z_{(1)}^{e^4}(v)$ and their derivatives. 
Repeated from \AT's Eqs.~(5.25), (5.26), and (5.27), they read
\begin{equation}
a(v) = \sqrt{1-3v}\,  z_{(1)}^{e^0}(v) -v \hh{1+ \frac{1-4v}{\sqrt{1-3v}}} , \label{eq:a_master}
\end{equation}
\begin{equation}
\label{eq:dBar_master}
\begin{aligned}
\bar{d}(v)&=
	\frac{
		v \hh{7 -\frac{141}{4}v+45 v^2}
	}{
		2 (1-3 v)^{5/2}
	}
	-
   	\frac{
   		\hh{1-\frac{21}{8}v} 
   	}{
   		(1-3 v)^{3/2}
   	}
   	z_{(1)}^{e^0}(v)
\\
	&\qquad+
	\frac{
		\hh{2-\frac{51}{2}v+101 v^2-132 v^3} 
	}{
		\sqrt{1-3 v}(1-6 v)^2 
	}z_{(1)}^{e^0}{}'(v)
 \\ 
   &\qquad
   -\frac{
   		v (1-2 v)\sqrt{1-3 v}   
   	}{
   		2 (1-6v)
   	}z_{(1)}^{e^0}{}''(v)
\\
	&\qquad
   	+
   	\frac{
   		(1-2 v)\sqrt{1-3 v} 
	}{
   		v(1-6 v) 
   	}z_{(1)}^{e^2}(v),   
\end{aligned}
\end{equation}
and
\begin{widetext}
\begin{equation}
\label{eq:q_master}
\begin{aligned}
 q(v) &=
 \frac{
 	9 v(1-2 v)^2 \hh{1-\frac{47}{9}v+\frac{1349 }{144}v^2-\frac{71}{12}v^3} 
 }{
 	8 (1-3v)^{7/2}
 }
 -
 \frac{
 	5 v (1-2 v)^2\hh{1-\frac{15}{8}v}  
 }{
 	16 (1-3 v)^{5/2}
 }
 z_{(1)}^{e^0}(v)
 \\
 &\quad+
 2 \frac{
 	(1-2 v)^2 
 	 \hh{
  1-\frac{100}{3}v +\frac{22963 }{48}v^2 -\frac{372085}{96}v^3 +\frac{467057}{24}v^4  -\frac{185935}{3}v^5 +\frac{243789}{2}v^6 -\frac{269793}{2}v^7 + 64188 v^8
 }
 }{
 	v (1-3 v)^{3/2} (1-6 v)^5  
 }
 z_{(1)}^{e^0}{}'(v)
 \\
 &\quad
 -
 \frac{
 	7 v (1-2v)^3\hh{1-\frac{285}{28}v-\frac{299}{14}v^2 + \frac{1851}{7}  v^3 -\frac{2790 }{7}v^4}
 }{
 	12  \sqrt{1-3 v}(1-6 v)^4
 }
 z_{(1)}^{e^0}{}''(v)
  -
 \frac{
 	 v(1-2 v)^4\sqrt{1-3 v} \hh{1-\frac{25}{2}v+24 v^2} 
 }{
 	6 (1-6 v)^3
 }
 z_{(1)}^{e^0}{}'''(v)
 \\
 & \quad
 + 
 \frac{
 	v^2 (1-2 v)^4 (1-3 v)^{3/2}   
 }{
 	24 (1-6 v)^2
 }
 z_{(1)}^{e^0}{}''''(v)
 -
 \frac{
 	7(1-2 v)^3\hh{1-\frac{99 }{4}v +\frac{3097}{14}v^2 - \frac{5214}{7} v^3 +828 v^4}  
 }{
 	6 v \sqrt{1-3 v} (1-6 v)^4  
 }
 z_{(1)}^{e^2}(v)
 \\
% &\quad
% -
% \frac{
% 	7(1-2 v)^3\hh{1-\frac{99 }{4}v +\frac{3097}{14}v^2 - \frac{5214}{7} v^3 +828 v^4}  
% }{
% 	6 v \sqrt{1-3 v} (1-6 v)^4  
% }
% z_{(1)}^{e^2}(v)
% \\
 & \quad
 +
 \frac{
 	(1-2v)^4\sqrt{1-3 v}\hh{1-\frac{15}{2}v}  \hh{1-\frac{8}{3}v}  
 }{
 	v(1-6 v)^3
 }
 z_{(1)}^{e^2}{}'(v) 
 -
 \frac{
 	(1-2 v)^4(1-3 v)^{3/2}  
 }{
 	6 (1-6 v)^2
 }
 z_{(1)}^{e^2}{}''(v)
% \\
% & \quad 
 +
 \frac{
 	(1-2 v)^4(1-3 v)^{3/2}  
 }{
 	9  v^2(1-6v)^2
 }
 z_{(1)}^{e^4}(v).
\end{aligned}
\end{equation}
\end{widetext}

\begin{table}[tbph]
    \caption{The notation for the various functions and partial derivatives displayed in Eqs.~(\ref{eq:dBar_master}) and (\ref{eq:q_master}). The first column lists our notation. The second and third columns list the corresponding notations in \AT~and \BDG.}\label{table:notation}  
     \centering
    \begin{tabular}{rcp{8.9em}|l|p{6em}}
        \toprule 
            \multicolumn{3}{c|}{Notation here}& In \AT & \multicolumn{1}{c}{In \BDG~($n$=1)}\T\B\\          
    \hline 
	$U(v,e)$			& &	\raggedright (generalized) redshift					& $\avg{U}(v,e)$			& $U(v,e)$	\T\B \\
	$U_{(n)}(v,e)$  	& & \raggedright $\mathcal{O}(\nu^n)$ part of $U(v,e)$		& $\avg{U}_{(n)}(v,e)$		& {\raggedright $U_0(v,e)$,  $\delta U(v,e)$\footnotemark[1]	\B}\\
	$U_{(n)}^{e^0}(v)$  & & \raggedright circular-orbit value of $U_{(n)}$ 		& $U_{(n)}(v)$				& $\delta U^{e^0}(v)$	\B\\
	$U_{(n)}^{e^k}(v)$  & & \raggedright $\Lim{e\to 0}\npd{U_{(n)}(v,e)}{e}{k}$ & $\avg{U}_{(n)}^{e^k}(v)$	& $\delta U^{e^k}(v)\times k!$	\B\\
	$z(v,e)$			& &	\raggedright $U(v,e)^{-1}$							& $\avg{z}(v,e)$			& $z_1(v,e)$\T\B \\
	$z_{(n)}(v,e)$  	& & \raggedright $\mathcal{O}(\nu^n)$ part of $z(v,e)$		& $\avg{z}_{(n)}(v,e)$		& $U_0(v,e)^{-1}$, $\delta z_1(v,e)$\footnotemark[1]	\B\\
	$z_{(n)}^{e^0}(v)$  & & \raggedright circular-orbit value of $z_{(n)}$ 		& $z_{(n)}(v)$				& $\delta z_1^{e^0}(v)$	\B\\
	$z_{(n)}^{e^k}(v)$  & & \raggedright $\Lim{e\to 0}\npd{z_{(n)}(v,e)}{e}{k}$ & $\avg{z}_{(n)}^{e^k}(v)$	& $\delta z_1^{e^k}(v)\times k!$	\B\\
    \botrule
    \end{tabular}
	\footnotetext[1]{\BDG~strictly use the subscript $0$ for background quantities and $\delta$ for $\ord(m_1/m_2)$ quantities.}
\end{table}

\subsubsection{Some remarks regarding notation and nomenclature}\label{sec:notation}
The use of terminology and notation for the redshift in the literature is far from standardized. We therefore take a moment to clarify the terminology and notation used in this paper. Depending on the literary source both $U$ and $z$ are referred to as the 
``redshift''. Following Ref.~\cite{Barack:2011ed}, we refer to the quantity $U$ defined in Eq.~\eqref{eq:defredshift} as the redshift.  Its reciprocal $z$ is referred to as the inverse redshift here. Note that this is the opposite terminology to the one used in \AT~and \BDG.

In Table~\ref{table:notation} we summarize the notation used in this paper for the various expansions of the (inverse) redshift. For comparison, we also include the notation used by \AT~and \BDG~for the same quantities. We note in particular that \BDG~absorb the factors of $1/2!, 1/4!$ 
into their inverse redshift quantities $\delta z_1^{e^2}(v), \delta z_1^{e^4}(v)$.
\section{Numerical methods}\label{sec:computation}
\subsection{Analytic expressions for \texorpdfstring{$z_{(1)}^{e^0}(v)$}{} and its derivatives}\label{sec:z1e0_and_derivatives}
Looking at Eqs.~(\ref{eq:dBar_master}) and (\ref{eq:q_master}) we see that we will need to take $v$ derivatives up to the fourth order for $z_{(1)}^{e^0}(v)$ and second order for $z_{(1)}^{e^2}(v)$.
We will provide details for the computation of the latter derivatives in Sec.~\ref{sec:q_computation}. For now, we focus on $z_{(1)}^{e^0}(v) $ and its $v$ derivatives. This quantity can be obtained 
in a straightforward manner from the GSF quantity $h^R_{uu} \equiv h^R_{\alpha\beta}\,u_{(0)}^\alpha u_{(0)}^\beta$ where $u^\alpha_{(0)}$ is the particle four-velocity. Ref.~\cite{Akcay_et_al2012} computed $h^R_{uu}$ hence $z_{(1)}^{e^0}(v)$ to a fractional accuracy of $\gtrsim 10^{-10}$. More recently, Ref.~\cite{Do.al.15} presented 18-digit accurate numerical data for $U_{(1)}^{e^0}(v)$. These approaches are based on directly solving the Einstein 
field equations for the metric perturbation in respective gauges of Lorenz and Regge-Wheeler.

On a parallel front, solving the Teukolsky equation for the Weyl scalar $\psi_4$ using the so-called Mano-Suzuki-Tagasuki (MST) method \cite{Ma.al.96}, expanding the resulting hypergeometric functions at $v=0$ then reconstructing the metric perturbation via Cohen-Chrzanowski-Kegeles (CCK) reconstruction 
\cite{Chrzanowski:1975wv,Cohen:1974cm,Kegeles:1979an,Wald:1978vm} have yielded very high-order pN expansions for $h^R_{uu}$ \cite{Shah:2013uya}.
Most recently, Ref.~\cite{Chris2015} obtained $U_{(1)}^{e^0}(v)$ to $\ord(v^{23.5})$ i.e. 21.5pN. However, even at such a high order, the pN series degrades quickly in the strong-field regime ($v \gtrsim 1/10$), as 
comparisons with the data of Refs.~\cite{Akcay_et_al2012} and \cite{Do.al.15} clearly show. As a result, we have opted to construct a 
`hybrid' $z_{(1)}^{e^0}(v)$ given by the following piecewise function
\be
z_{(1)}^{e^0}(v) = \left\{\begin{array}{l}z_{(1)}^{e^0,Pad\acute{e}}(v),\quad v\ge v_c \T\B \\ z_{(1)}^{e^0,23.5}(v),\ \quad v < v_c \T\B\end{array}\right. , \label{eq:z1_piecewise}
\ee
where $z_{(1)}^{e^0,23.5}(v)$ is obtained via Eq.~(\ref{eq:z1circ}) using  the expression for $U_{(1)}^{e^0,23.5}(v)$ from Ref.~\cite{Chris2015} and $z_{(1)}^{e^0,Pad\acute{e}}(v)$ is obtained from a Pad\'{e} fit to the strong-field $U_{(1)}^{e^0}(v)$ results of Ref.~\cite{Do.al.15} 
where we have picked their data 
in the range $v\in[1/30,1/5]$ and supplemented it with a few more points near $v=1/20$ using $z_{(1)}^{e^0,23.5}(v)$, which agrees with all of Ref.~\cite{Do.al.15}'s digits for $v<1/20$. We construct Pad\'{e} fits to this data set of the form
\be
U_{(1)}^{e^0,Pad\acute{e}}(k,n,v)\equiv \frac{v\;(1+\sum_{i=1}^k A_i v^i)\B}{(1-3v)^2(1+\sum_{j=1}^n B_j v^j)\T}, \label{eq:Pade_form}
\ee
where $A_i, B_j$ are the fitting coefficients and $k \le n$. The factor of $v/(1-3v)^2$ represents the leading order $v\to 0$ and $v\to 1/3$ behaviors. These were extracted from the work of
Ref.~\cite{Akcay_et_al2012} via Eq.~(\ref{eq:a_master}) above.
We have experimented with various Pad\'{e} fits such that $k+n <$ (\# data points). We have checked the faithfulness of the fits by comparing how well they approximate the unused data as well as how well they match the 21.5pN expression for $x\le 1/20$.
For our final result, we have settled on
\be
z_{(1)}^{e^0,Pad\acute{e}}(v) = -\frac{U_{(1)}^{e^0,Pad\acute{e}}(7,8,v)}{U_{(0)}^2(v)}
\ee
which matches the data to $ \lesssim 10^{-15}$. We performed a further check of our fit and its first and second derivatives by constructing $a(v)$ via Eq.~(\ref{eq:a_master})
%
%\be
%a(v) = \sqrt{1-3v}\: z_{(1)}^{e^0,Pad\acute{e}}(v) - v\left( 1+ \frac{1-4v}{\sqrt{1-3v}}\right) \label{eq:EOB_a_potential}
%\ee
%
and evaluating $\{a(1/6), a'(1/6), a''(1/6)\}$ to compare these quantities with those of Ref.~\cite{Akcay_et_al2012} which were computed to high accuracy. % using local fits around $v=1/6$ and ninth-order finite-difference derivatives. 
%Although we will not actually compute $\bar{d}(v)$ and $q(v)$ at $v=1/6$ ($v=1/6$ is an unstable point) 
%we nonetheless 
%We use this information at $v=1/6$ as a worst-case check of our Pad\'{e} fit for $z_{(1)}^{e^0}(v)$ and its derivatives. 
We find that our fit yields values for $\{a(1/6), a'(1/6), a''(1/6)\}$ that agree with Ref.~\cite{Akcay_et_al2012} to 
$\{ \sim10^{-10}, < 10^{-8}, < 10^{-8}\}$. As there is no available data to perform a similar check for $z_{(1)}^{e^0,Pad\acute{e}}\,{}'''(v)$ and $ z_{(1)}^{e^0,Pad\acute{e}}\,{}''''(v)$ we make do with computing the error for these quantities using the 
standard methods which we also employ to compute the errors in $z_{(1)}^{e^0,Pad\acute{e}}(v),\, z_{(1)}^{e^0,Pad\acute{e}}\,{}'(v),\, z_{(1)}^{e^0,Pad\acute{e}}\,{}''(v)$.

The matching point in Eq.~(\ref{eq:z1_piecewise}), $v_c$, is determined empirically by numerically evaluating the largest-order pN term at a value of $v$ such that its magnitude is 
$\lesssim 5\times 10^{-12}$. For $z_{(1)}^{e^0}(v)$ this gives $v_c\approx 1/10$. As the unknown higher-order pN terms at $v=1/10$ would most likely yield a number larger than $5\times 10^{-12}$,  
we expect that the known pN expressions should have an absolute accuracy of about $10^{-10}$  at $v=v_c$. We confirm this accuracy estimation by explicitly computing the relative difference between $U_{(1)}^{e^0,23.5}(v)$ and the numerical data of Ref.~\cite{Do.al.15} 
in the vicinity of the matching. %We similarly construct $\left\{z_{(1)}^{e^0}'(v), z_{(1)}^{e^0}''(v), z_{(1)}^{e^0}'''(v), z_{(1)}^{e^0}''''(v)  \right\}$ by analytically differentiating $z_{(1)}^{e^0}^{Pad\acute{e}}(v),\, z_{(1)}^{e^0}^{23.5pN}(v)$ 
%and empirically determining the ideal position of the cut-off. 
We move $v_c$ to smaller values as we take higher-order derivatives since the pN series loses a power of $v$ with each differentiation. At each derivative order, we determine $v_c$ anew using the aforementioned empirical method. By the fourth derivative, 
$v_c$ moves out to $1/25$.

With $v_c$ determined at each derivative order we construct the derivatives of $z_{(1)}^{e^0}(v)$ as piecewise functions by analytical differentiations of $z_{(1)}^{e^0,23.5}(v)$ and $ z_{(1)}^{e^0,Pad\acute{e}}(v)$. This naturally introduces a discontinuity to 
each derivative at the corresponding $v_c$. We have checked that the size of these jumps relative to the magnitude of the derivatives is small (ranging from $\sim 10^{-12}$ for first derivative to $\sim 10^{-8}$ for the fourth). We further make sure to exclude all $v_c$'s 
from our $v$ grid for the data sets. %We hope to improve on our current model once more strong-field high-accuracy data for $U_{(1)}^{e^0}$ becomes available.
\subsection{Computation of \texorpdfstring{$ z_{(1)}^{e^2}(v)$}{} and \texorpdfstring{$ z_{(1)}^{e^4}(v)$}{}}\label{sec:U1e2_computation}
To compute $ z_{(1)}^{e^2}(v)$ and $  z_{(1)}^{e^4}(v)$ we use Eqs.~(\ref{eq:z1e2}) - (\ref{eq:Lim_e0_d4Ude4}) where we compute $ U_{(1)}^{e^2}(v)$ and $  U_{(1)}^{e^4}(v)$ by fitting polynomials in powers of $e^2$ to the numerical data for $U_{(1)}(v,e)$ at each $v$. 
We use two independent approaches: (i) fitting polynomials directly to $U_{(1)}(v,e)$ data obtained from the \verb|C|-based Lorenz-gauge code of Refs.~\cite{Akcay:2015pza,Ak.al.13}, (ii) using the \verb|Mathematica|-based Teukolsky-MST-CCK code of 
Ref.~\cite{Maarten2015} which extracts the $e^0, e^2$ and $e^4$ dependence of the numerically computed multipole $l$ modes of $h^R_{uu}$ then constructs power-law fits to the resulting three separate sets of mode data. The large-$l$ modes fall off as power-law `tails' whose behavior 
is well understood, since the work of Barack in Ref.~\cite{Barack:1999wf}, and was thoroughly studied in Ref.~\cite{Heffernan_et_al2014}. As approach (i) is limited to machine precision, the resulting data has an accuracy of $\gtrsim 10^{-11}$. On the other hand, 
approach (ii) uses \verb|Mathematica|'s arbitrary precision algorithms so in principle $U _{(1)}$ can be obtained to arbitrarily high accuracies albeit with increasing computational burden. We find that our respective codes agree to 
$\sim 10^{-10}$ for $v\in [1/75, 1/9]$ and slightly less at the edges of the $v$ space due to the limitations of the Lorenz-gauge code (cf. \cite{Ak.al.13}). As approach (ii) is more accurate we use its results for our final values presented in Sec.~\ref{sec:results} 
and use the Lorenz-gauge code to check these as best as we can.

We compute $U_{(1)}(v,e)$ along eccentric orbits over an evenly spaced grid in the $(v,e)$ parameter space where $v$ ranges from $1/1200$ to $199/1200$ with grid spacing of $1/1200$. Since we are interested in extracting only the $\ord(e^2)$ and $\ord(e^4)$ 
contributions to $U_{(1)}$ we focus on small eccentricities which, for the Lorenz-gauge code, range from 1/200 to 1/20 with grid spacing of $1/200$. Approach~(ii) uses smaller eccentricities as explained below. 
We also add the circular-orbit result $U_{(1)}^{e^0}$ to our eccentricity data set at each $v$ value. We further make use of the fact that 
$U_{(1)} \sim -v + \ord(v^2) \to 0 $ as $v\to 0$, which gives us a `free' point to add to our data sets at $v=0$. We now provide more details for each approach.
\subsubsection{The Lorenz-gauge based method} \label{sec:Lorenz_gauge_method}
We use Lorenz-gauge data only for $ 1/75 \le v \le 3/20$ with a relative error of $ 10^{-10}$ for $v \lesssim 1/9$ increasing to $ 10^{-8}$ at $v=3/20$.
The details of the computation for $U_{(1)}(v,e)$ are thoroughly explained in Refs.~\cite{Akcay:2015pza}, \cite{Ak.al.13} so here, we focus on the fitting procedure. 
We used the following four polynomials in $e^2$ for our fits:
\begin{align}
 &\begin{aligned}
 	\mathrm{fit}_1 &= a + b\, e^2 + c\, e^4, 
 	\\
 	\mathrm{fit}_2 &= \mathrm{fit}_1 + d \,e^6
 \end{aligned}
 \text{(without the $e=0$ data)}, \label{eq:poly_fits12}
 \\
 &\begin{aligned}
	 \mathrm{fit}_3 &=b\, e^2 + c\, e^4 
	\\	
	 \mathrm{fit}_4 &= \mathrm{fit}_3 + d \,e^6                     
 \end{aligned}
 \qquad\text{(with $e=0$ data)}, \label{eq:poly_fits34}
\end{align}
These yield two values for $a$: $\{a_1,a_2\}$ and four for $b$ and $c$: $\{b_1,\ldots,b_4\},\ \{c_1, \ldots c_4 \}$. Although from Eq.~(\ref{eq:U_e2_expansion}) we have that $a= U_{(1)}^{e^0}$, 
we do not use this information for fit$_1$ and fit$_2$ so that we can perform two checks of the fits immediately by defining an average $\bar{a}\equiv (a_1+a_2)/2$ and 
a fit error $\Delta a \equiv \max\{| a_2-a_1|, \Delta a_1,\Delta a_2\}$ where $\Delta a_{1,2}$ are absolute errors for $a_{1,2}$ obtained from linear regression methods used for fit$_{1,2}$.
For our first check we compute the relative difference between $ \bar{a}$ and the true result $U_{(1)}^{e^0}$ and find this to be $\lesssim 10^{-10}$ for $v \lesssim 1/7$. This decreases by a few more orders of magnitude as $v\to 0$. 
Then we check that $|U_{(1)}^{e^0}-\bar{a}| \le \Delta a$ for all $v$ consistent with our expectation that the true result should lay within the error region of the approximation from the fits.

Similarly, we construct $\bar{b}$ from the average of $\{b_1,b_2,b_3,b_4\}$ and its error $\Delta b$ from $\max\{| \bar{b}-b_i |, \Delta b_i \} $ with $i=1,\ldots, 4$. We obtain $\bar{c}$ and $\Delta c$ in an analogous fashion. Our error estimation ensures that we retain only the significant digits for $\bar{a}, \bar{b},\bar{c}$ agreed upon by all four fits (two for $\bar{a}$). We compute the errors for $ z_{(1)}^{e^2}(v)$ and $z_{(1)}^{e^4}(v)$ 
by adding $\Delta\{a,b,c\}$ in quadrature using Eqs.~\eqref{eq:z1e2} and \eqref{eq:z1e4} while taking into account the fact that the errors are correlated hence the resulting covariance matrix has off-diagonal elements.
\subsubsection{The Teukolsky-MST-CCK method}\label{sec:TMSTCCK_method}
In a recent paper \cite{Maarten2015}, one of us presented a method for calculating $\la h_{uu}^R\ra$ for eccentric orbits using the radiation-gauge techniques pioneered by Friedman \emph{et al.} \cite{Keidl:2006wk,Keidl:2010pm,Shah:2010bi,Shah:2012gu}. 
Like the Lorenz code above this method is based on a frequency-domain decomposition and the method of extended homogeneous solutions. However, instead of solving a coupled set of equations to find the Lorenz-gauge metric perturbation directly, 
the method first solves the Teukolsky equation to determine the Weyl scalar $\psi_4$. The retarded metric perturbation is then obtained in the (outgoing) radiation gauge by inverting the differential operator for $\psi_4$ using the formalism of 
Chrzanowski, Cohen, Kegeles, and Wald \cite{Chrzanowski:1975wv,Cohen:1974cm,Kegeles:1979an,Wald:1978vm}. Since this operator is not injective, this inversion is ambiguous up to an element of its kernel. The gauge-invariant content of this kernel is simply given 
by a shift in mass and angular momentum of the system \cite{Wald:1973}, and can be extracted unambiguously \cite{MOPVM}. As shown in Ref.~\cite{Pound:2013faa}, the regular metric perturbation can then be obtained using a mode-sum regularization scheme.

Since the whole method is implemented using arbitrary-precision arithmetic and uses a numerical implementation \cite{Fujita:2004rb,Fujita:2009bp,Meent:2015a} of the analytical series solution to the Teukolsky equation devised by Mano, Suzuki, 
and Takasugi \cite{Mano:1996gn,Mano:1996vt}, individual modes can be solved to almost any desired accuracy. The limiting step in the accuracy of this method comes from fitting for the large-$l$ tail of the mode sum. Although we have faster-than-polynomial 
convergence in the number of $l$ modes, the convergence of this sum is known \cite{Akcay_et_al2012} to be slow in the strong-field regime. Since computing more $l$ modes\footnote{Note that the restriction of our implementation to modes with $l\leq 30$ originally 
reported in Ref.~\cite{Maarten2015}, has since been resolved allowing calculation of any $l$ mode, given sufficient time.} is very time consuming this limits the accuracy that can be achieved in the strong field with this method to slightly more than the Lorenz-gauge code.

To take full advantage of the highly accurate mode calculations of this method, we adopt an alternative approach to obtain the $e^2$ expansion of $U_{(1)}(v,e)$. Using the Teukolsky-MST-CCK code we calculate the individual regularized $l$ modes $\la h^{R,l}_{uu}\ra(v,e)$ 
to a relative accuracy of $10^{-25}$ for a range of orbits with the same value of $v$ and varying eccentricity $e$. We then extract the expansion of $\la h^{R,l}_{uu}\ra(v,e)$ in $e^2$ as before using fits of the form \eqref{eq:poly_fits34}, 
obtaining $\la h^{R,l}_{uu}\ra^{e^2}(v)$ and $\la h^{R,l}_{uu}\ra^{e^4}(v)$. Assuming the order of the $e \to 0$ and $l\to\infty$ limits can be exchanged, the $e^2$ expansion coefficients of  $\la h^{R}_{uu}\ra(v,e)$ are now given by
\begin{align}
\la h^{R}_{uu}\ra^{e^2}(v) &= \sum_{l=0}^{\infty} \la h^{R,l}_{uu}\ra^{e^2}(v),\quad\text{and}\\
\la h^{R}_{uu}\ra^{e^4}(v) &= \sum_{l=0}^{\infty} \la h^{R,l}_{uu}\ra^{e^4}(v),
\end{align}
where the infinite sums over $l$ are to be performed as is usual in self-force calculations by calculating the partial sums and estimating the remaining `large-$l$ tail' by fitting a power series in $l^{-1}$. The expansion coefficients of $U_{(1)}$ are finally 
obtained from%by using that 
\be
U_{(1)} = \frac{1}{2}U_{(0)} \la h_{uu}^R\ra . 
\ee
%By further capitalizing on the accuracy of the mode calculations in the Teukolsky-MST-CCK method by using much smaller eccentricities ranging between $10^{-6}$ and $10^{-2}$, we are able to calculate $U_{(1)}^{e^2}$ and $U_{(1)}^{e^4}$ at accuracies of $10^{-10}$ and $10^{-6}$ respectively at $v=49/300$ dropping to $10^{-17}$ and $10^{-14}$ at $v=1/300$.
Thanks to the high accuracy of the mode calculations in the Teukolsky-MST-CCK method, we are able to calculate $U_{(1)}^{e^2}$ and $U_{(1)}^{e^4}$ at  $v=199/1200$ to accuracies of $\sim 10^{-10}$ and $\sim 10^{-6}$, respectively by using much smaller eccentricities ranging between $10^{-6}$ and $10^{-2}$. By $v=1/1200$, the accuracies improve to $\sim 10^{-22}$ and $\sim 10^{-18}$, respectively.

To confirm that this procedure of switching the order of the $l$ summation and the $e$ fitting works, we compared the resulting values for the $U_{(1)}^{e^2}$ and  $U_{(1)}^{e^4}$ with the same coefficients obtained using the more traditional procedural order 
applied to the results from the Lorenz-gauge \verb|C|-code. These match the found coefficients to within their (obviously larger) errors.
\subsection{Computation of \texorpdfstring{$ z_{(1)}^{e^2}{}'(v)$}{} and \texorpdfstring{$ z_{(1)}^{e^2}{}''(v)$}{}}\label{sec:z1e2_prime_computation}%
As can be seen from Eq.~(\ref{eq:q_master}), we need to compute first and second $v$ derivatives of $z_{(1)}^{e^2}(v)$. As we have a large data set with 200 elements (including $v=0$) with a grid spacing of $h=1/1200$, we use finite differencing (FD) to compute the derivatives. Due to the fact that $z_{(1)}^{e^2} \sim (1-6v)^{-1} $ 
as $v\to 1/6$ we choose to compute the FD derivatives for the rescaled function $ \tilde{z}_{(1)}^{e^2}\equiv (1-6v) z_{(1)}^{e^2}$. We find that this significantly improves our results for the FD derivatives near $v=1/6$.
%Similarly, we define $ \tilde{z}_{(1)}^{e^4}\equiv (1-6v)^3 z_{(1)}^{e^2}$ and compute its FD derivatives in our computation for $q(v)$.
This singular behavior of $z_{(1)}^{e^2}(v)$ along with that of $ z_{(1)}^{e^4}(v)$ have been studied by \BDG; we provide our own analysis in Sec.~\ref{sec:Behavior_z1e2_z1e4}.

We compute the derivatives $\tilde{z}_{(1)}^{e^2}{}'(v), \tilde{z}_{(1)}^{e^2}{}''(v)$ at FD orders ranging from five to nine and check that the derivatives converge as the FD order is increased. Since the data has limited accuracy, finite differencing `saturates' once the grid resolution error is comparable to the errors in the data.
Our analyses show that we hit this saturation bound at a FD order of $\sim 9$. So in general, we do not go beyond ninth order FD derivatives. The convergence of the FD derivative for $ \tilde{z}_{(1)}^{e^2}{}'$ ranges from 
$\sim 10^{-14}$ near $v=0$ to $\sim 10^{-8}$ near $v=1/6$. Similarly, for $ \tilde{z}_{(1)}^{e^2}{}''$, the convergence ranges from $\sim 10^{-12}$ to $\sim 10^{-5}$. As we reach the edges of our numerical grid (i.e. $v=0,\,1/6$), the FD derivatives suffer from the usual edge effects so the convergence naturally jumps up by a 
few orders of magnitude.

To actually compute the FD derivatives we use \verb|Mathematica|'s \verb|NDSolve`FiniteDifferenceDerivative| function. We compute the errors for each derivative by using the corresponding stencil formula in the standard quadrature error computation. As the error for each  grid point is obtained independently from the others, the errors 
are not correlated. Our routine readily works for any derivative order and any stencil from edge points to midpoints. Our estimated errors for $\tilde{z}_{(1)}^{e^2}{}'(v)$ range from $\sim 10^{-14}$ near $v=0$ to $\sim 10^{-8}$ near $v=1/6$. Similarly, the errors for $\tilde{z}_{(1)}^{e^2}{}''(v)$ vary from $\sim 10^{-11}$  
to $\sim 10^{-5}$.
\section{Results}\label{sec:results}
\subsection{Behavior of \texorpdfstring{$z_{(1)}^{e^2}$}{z1e2} and \texorpdfstring{$z_{(1)}^{e^4}$}{z1e2}}\label{sec:Behavior_z1e2_z1e4}
%\begin{figure}[t]
%\includegraphics[width=0.99\linewidth]{Figs/z1e2_Root_Plot.pdf}%
%  \caption{Our numerical data for $z_{(1)}^{e^2}(v)$ and the fit to it showing that it changes sign at $p=1/v\approx 6.7598$.}
%\label{fig:z1e2_Root_Plot}
%\end{figure}
As discussed in \BDG, the function $z_{(1)}^{e^2}(v)$ becomes singular as it approaches the innermost stable circular orbit (ISCO) at $v=1/6$. Our data confirms that $z_{(1)}^{e^2}(v)$ has a simple pole at $v=1/6$. Moreover, we are able to numerically extract the first few terms of its Laurent expansion,
\begin{equation}\label{eq:ze2nearISCO}
%\begin{split}
 z_{(1)}^{e^2} = \sum_{i=-1}^\infty c_i^{e^2} (1-6v)^i,
%\end{split}
\end{equation}
with
\begin{equation}
\begin{split}
 c_{-1}^{e^2} &= +0.01364554556(2),\\
 c_{0}^{e^2} &= -0.116733823(2),\\
 c_{1}^{e^2} &= -0.0910091(4),\\
 c_{2}^{e^2} &= +0.519971(2),\\
 c_{3}^{e^2} &= -0.8245(3), \\
 c_{4}^{e^2} &= +1.1503(5),\\
 c_{5}^{e^2} &= -1.45(2),\quad\text{and}\\
 c_{6}^{e^2} &= +1.9(4),
\end{split}
\end{equation}
where the number in parentheses indicates the approximate error. Based on older self-force data \BDG~provide the estimates  $c_{-1}^{e^2} \approx 0.0136455$ and $c_{0}^{e^2} \approx -0.116733$, which fully agree with our values. They also correctly conclude that since  $z_{(1)}^{e^2}(v)$ is negative in the weak-field limit, 
it must change sign (at least once) between $v=0$ and $v=1/6$. They estimate that this happens at $p=1/v\approx 6.760$. Our data yields $6.759785(2)$. 

The analysis of \BDG~also indicates that $z_{(1)}^{e^4}(v)$ has a third order pole at $v=1/6$. Our data again confirms this conclusion giving the following Laurent expansion,
\begin{equation}\label{eq:ze4nearISCO}
\begin{split}
 z_{(1)}^{e^4} = \sum_{i=-3}^\infty c_i^{e^4} (1-6v)^i,
\end{split}
\end{equation}
with coefficients together with the estimates provided by \BDG.
\begin{equation*}
\begin{array}{r|d{20}|d{10}}
			& \multicolumn{1}{c|}{\text{Here}}			
									&\multicolumn{1}{c}{\text{\BDG}}\\
\hline
 {c}_{-3}^{e^4} 	& +0.000426423298976(4)	&+0.0004263\T\\
 {c}_{-2}^{e^4} 	& -0.00127926989693(1)	&-0.001279\\
 {c}_{-1}^{e^4} 	& +0.00073197(1)		&+0.0006447\\
 {c}_{ 0}^{e^4} 	& -0.0942532(3)			&-0.09396\\
 {c}_{ 1}^{e^4} 	& +0.30778(7)			&+0.3435\\
 {c}_{ 2}^{e^4}		& -0.315(3)				&
\end{array}
\end{equation*}
The first two estimates of \BDG~appear to be spot on. However, their estimates for ${c'}_{\ge-1}$ significantly differ from our extracted values. This could be due to the lack of high-accuracy data available to \BDG~at the time of their computation. Finally, let us add that $z_{(1)}^{e^4}(v)$ also changes sign in the interval $v \in [0, 1/6]$. 
This happens approximately at $v = 0.1391647400(1)$. We are able to obtain more significant digits for this approximation compared with the sign change of $z_{(1)}^{e^2}(v)$ because the sign change happens farther away from the ISCO.

\subsection{The potential \texorpdfstring{$\bar{d}(v)$}{dbar}}\label{sec:d_bar_computation}
\begin{figure}[tb]
  \includegraphics[width=0.99\linewidth]{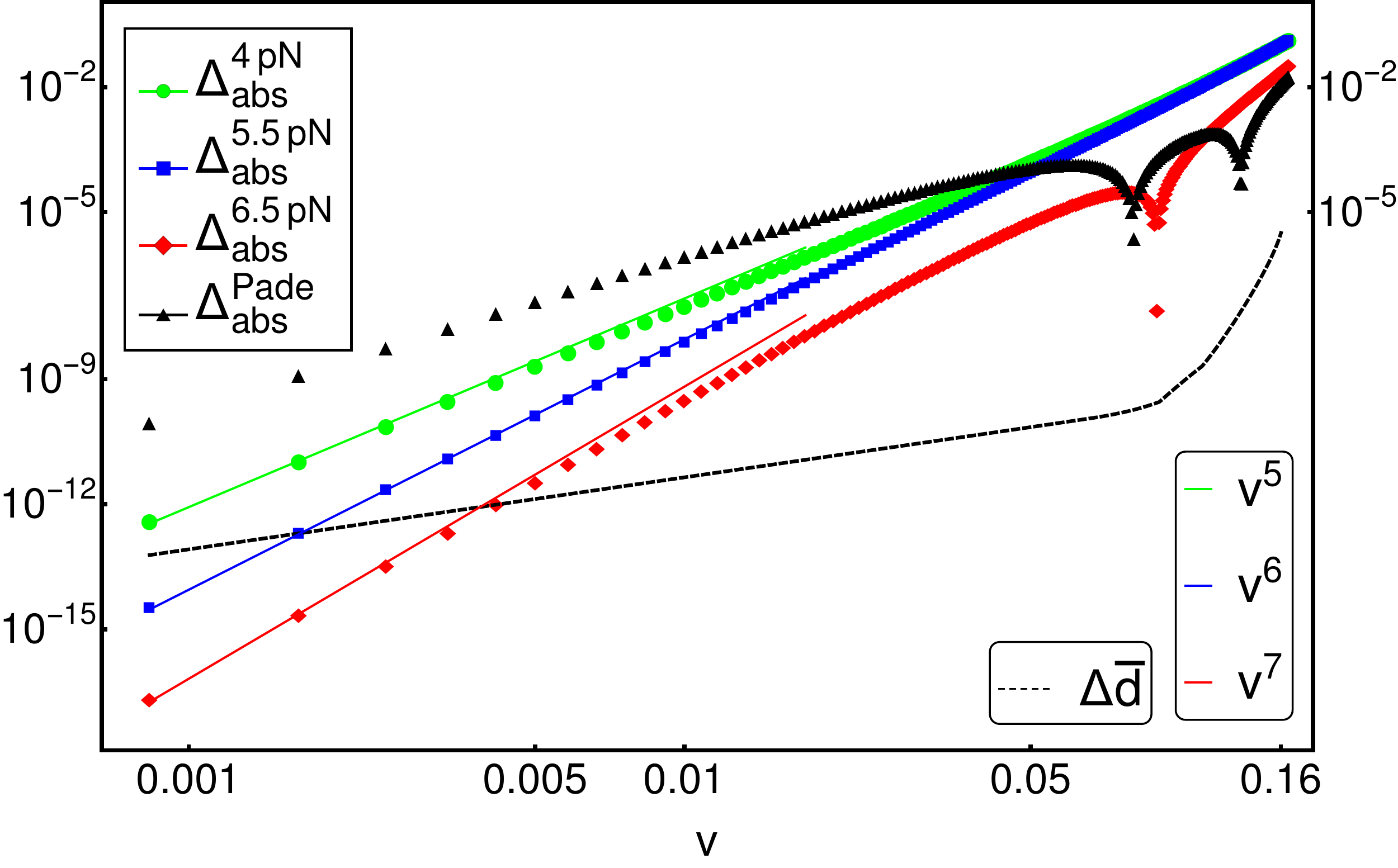}%
\caption{Absolute differences in $\bar{d}(v)$ between our numerical data and \BDG's corresponding pN series. The green dots, blue squares and red diamonds represent the difference between our data and the 4, 5.5, 6.5 pN expressions, respectively. The corresponding green, blue, red lines are fits that scale as $v^5, v^6, v^7$, respectively. 
The black triangles show the difference between our data and \BDG's Pad\'{e} fit. The dashed black curve ($\Delta\bar{d}$) is our estimated absolute error for $\bar{d}(v)$.}
\label{fig:d_bar_AbsDiff_vs_PN}
\end{figure}
\begin{figure}[tb]
  \includegraphics[width=0.99\linewidth]{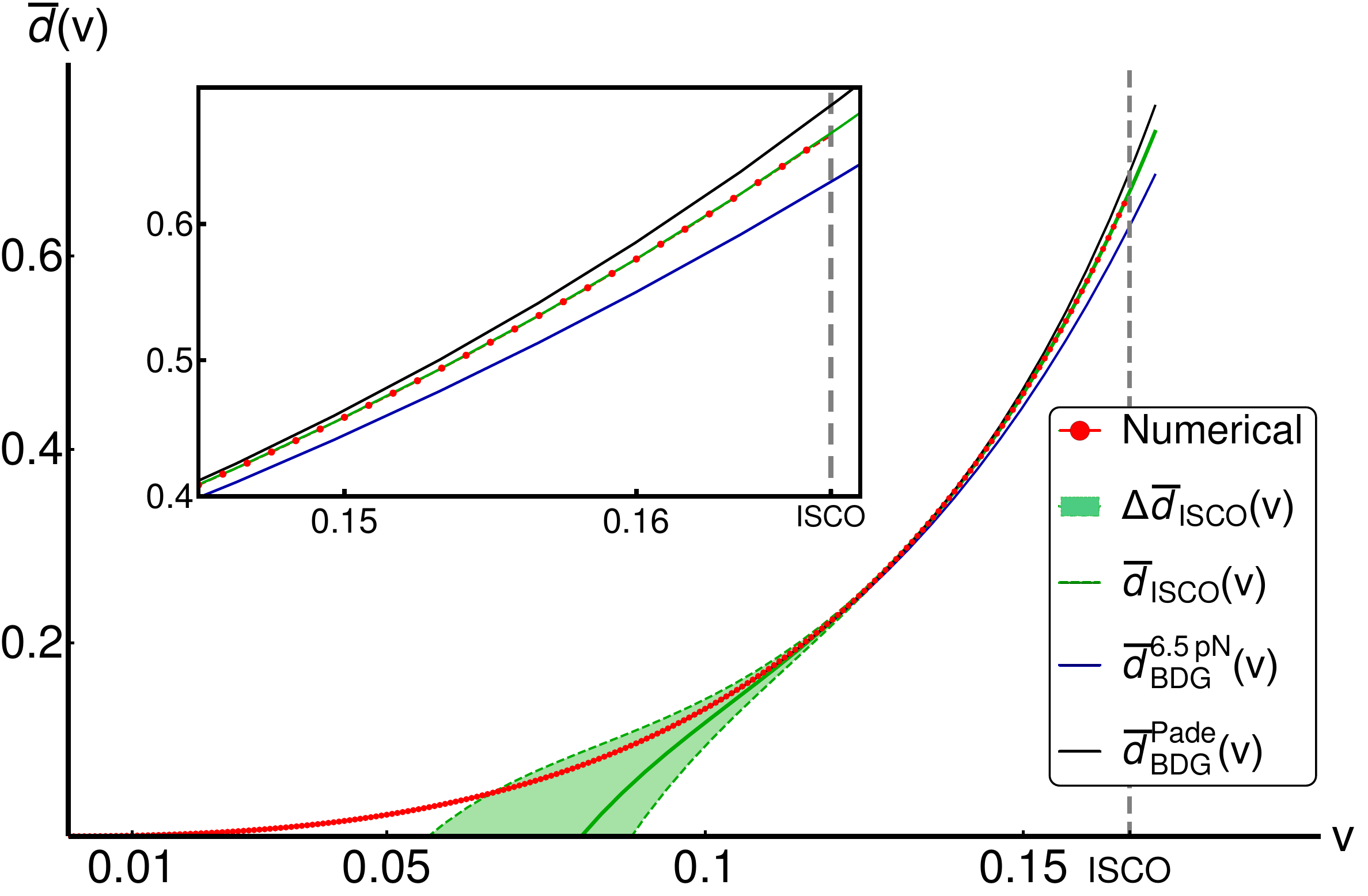}%
	\caption{Our numerical data for $\bar{d}(v)$ compared with \BDG's 6.5pN expression and their Pad\'{e} fit. The solid green curve given by Eqs.~\eqref{eq:dnearISCO} and \eqref{eq:dBars_coefficents} is our ISCO expansion for $\bar{d}(v)$. 
	The shaded green region ($\Delta \bar{d}_{\mathsf{ISCO}}(v)$) bounded by the green dashed lines represents our estimated error for the ISCO expansion.}
\label{fig:dbar_overview}
\end{figure}
\begin{table*}[ht]
    \caption{Numerical values of $\bar{d}(v)$ at various $v$ which overlap with those given by Refs.~\cite{Akcay_et_al2012} and \cite{Nori_pvt}. As can be seen, the recomputed values of $\bar{d}(v)$ by Ref.~\cite{Nori_pvt} are more consistent with our findings. 
	}
    \label{table:dbar_Nori_comparison}
\centering
	\begin{tabular}{@{}l| llllll@{}}
	    \toprule
	p=1/v \hspace{1.0cm} & \hspace{0.4cm}6.5 & \hspace{0.4cm}7& \hspace{0.4cm}7.5 &\hspace{0.4cm}8 & \hspace{0.4cm}8.5 & \hspace{0.4cm}9  \T\B \\ 
\hline
	$\bar{d}$(Ref.~\cite{Akcay_et_al2012}) & 0.5024(2)&0.38986(8) & 0.31129(5)& 0.25423(3) & 0.21141(2) & 0.17849(2)     \T\B \\ %  
	$\bar{d}$(Ref.~\cite{Nori_pvt}) & 0.4994(8) & 0.3884(4) & 0.3105(3) & 0.2537(1) & 0.2111(1) & 0.17828(6)     \T\B \\ 
	$\bar{d}$(Here) & 0.499909(1)&0.3886784(2) & 0.31066197(5)& 0.25382891(1) & 0.211156568(5) & 0.178312913(2)     \T\B \\ 
\hline
	p=1/v &\hspace{0.4cm} 9.5 & \hspace{0.4cm}10 & \hspace{0.4cm}11 & \hspace{0.4cm}12 &\hspace{0.4cm}13 & \hspace{0.4cm}13.5 \T\B \\
\hline
	$\bar{d}$(Ref.~\cite{Akcay_et_al2012}) & 0.15267(1) & 0.131940(9) & 0.101369(7) &0.080229(5) &  0.065016(5) & 0.058966(6)\T\B \\ %  
	$\bar{d}$(Ref.~\cite{Nori_pvt}) & 0.15247(5) & 0.13184(5) & 0.10131(2)  & 0.08017(4) & 0.06498(2) & 0.05894(4) \T\B \\ %  \T\B \\
	$\bar{d}$(Here)   & 0.152504936(1) & 0.1318652241(8) & 0.1013181313(3) &0.0801888618(2) &  0.0649853702(2)& 0.0589402300(2) \T\B \\
\hline
	p=1/v & \hspace{0.4cm}14 & \hspace{0.4cm}15 & \hspace{0.4cm}16 & \hspace{0.4cm}18 & \hspace{0.4cm}20 & \T\B \\
\hline
	$\bar{d}$(Ref.~\cite{Akcay_et_al2012}) & 0.053718(4)& 0.045101(6) & 0.038386(2)& 0.028753(3) & 0.0223171(7) &\T\B \\
	$\bar{d}$(Ref.~\cite{Nori_pvt}) &   0.05370(4) & 0.04509(3) & 0.03837(3) & 0.02874(4) & 0.02230(5)& \T\B \\ % 
	$\bar{d}$(Here)   & 0.0536924796(1) & 0.0450819583(1) & 0.0383711278(1) &0.02874267513(8) &  0.0223099574(7)& \T\B \\
    \botrule
    \end{tabular}
\end{table*}
From  $z_{(1)}^{e^0}(v)$ and  $z_{(1)}^{e^2}(v)$ we calculate $\bar{d}(v)$ using Eq.~(\ref{eq:dBar_master}). \BDG~have provided a pN series expansion for $\bar{d}(v)$ up to and including $\ord(v^{13/2})$. In Fig.~\ref{fig:d_bar_AbsDiff_vs_PN} we compare how well our numerical data matches their expression at several different pN orders. At each pN order the power-law decay of the residual towards $v=0$ is consistent with a term of the next pN order. 
In the weak field power-law behavior of the 6.5 pN residual even continues when the residual is much smaller than our estimated error. This indicates that our error estimate on $\bar{d}(v)$ in the weak field is too conservative. This is probably  due to an overly conservative error estimate on $z_{(1)}^{e^0}{}''(v)$.

The 6.5 pN residuals in Fig.~\ref{fig:d_bar_AbsDiff_vs_PN} also show that even in the weak field the residuals have not fully settled into their asymptotic $v^7$ behavior. This is indicative of the residuals not clearly separating into different higher-order pN terms. %separating contributions of different higher pN terms. 
Consequently, although we can visually identify the $v^7$ behavior, we do not expect to be able to numerically extract the missing 7 pN coefficients. Indeed, attempting to do so using the procedure described in Refs.~\cite{Akcay:2015pza} and \cite{Maarten2015}, i.e. by fitting $v^7$, $v^7\log v$, $v^7\log^2 v$ and higher order terms to the residual, yields inconclusive values for the coefficients of the fitting functions, which are the unknown higher-order pN parameters.

%{\bf Although it is difficult to tell from Fig. 3, the 6.5 pN residuals do not actually fully settle into their asymptotic} $v^7$ behavior. {\bf We confirmed this by using the procedure described in Refs.~\cite{Akcay:2015pza} and \cite{Maarten2015} by fitting} $v^7$, $v^7\log v$, $v^7\log^2 v$ and higher pN terms to the residual.
%{\bf This fitting yielded inconsistent values for the missing parameters. Consequently, we do not expect to be able to numerically extract the missing 7 pN coefficients. }

%The 6.5 pN residuals in Fig.~\ref{fig:d_bar_AbsDiff_vs_PN} also clearly show that even in the weak field the residuals have not fully settled into their asymptotic $v^7$ behavior. This is indicative of the residuals not clearly separating into different higher-order pN terms. %separating contributions of different higher pN terms. 
%Consequently, we do not expect to be able to numerically extract the missing 7 pN coefficients. Indeed, in attempting to do so using the procedure described in Refs.~\cite{Akcay:2015pza} and \cite{Maarten2015} by fitting $v^7$, $v^7\log v$, $v^7\log^2 v$ and higher pN terms yields inconclusive values for the missing parameters. 
% 
In Fig.~\ref{fig:d_bar_AbsDiff_vs_PN}, we also include \BDG's Pad\'{e} fit  (black triangles), which shows the best agreement with our numerical data in the strong-field regime as can be expected. The Pad\'{e} fit matches our data to better than 1\% for $v \le 3/20$. This difference is only slightly above 1\% beyond $v=3/20$.

In the strong-field regime, numerical data for $\bar{d}(v)$ has been presented in Refs.~\cite{Akcay_et_al2012,Ba.al.10}. Our comparisons with these data sets initially yielded a disagreement which was larger than their estimated errors (our errors are a few orders 
of magnitude smaller). More recently, Ref.~\cite{Nori_pvt} recomputed $\bar{d}(v)$ using the time-domain method of Ref.~\cite{Ba.al.10}, but this time to a higher maximum value for the multipole $l$ hence reducing the contribution of the large-$l$-tail fit to the overall 
result. In Table~\ref{table:dbar_Nori_comparison} we present a small subset of our numerical data for $\bar{d}(v)$ which overlaps that of Refs.~\cite{Nori_pvt,Akcay:2015pza}. The numerical data in the table shows that the recomputed values of $\bar{d}(v)$ are more 
consistent with ours. The recomputed error bars are larger than the previous estimations as the new results of Ref.~\cite{Nori_pvt} are preliminary. We expect these to decrease by one or two orders of magnitude once the recomputed results are finalized. 
On the other hand, our estimated errors for $\bar{d}(v)$ are much smaller partly due to the fact that our computation only needs the metric perturbation unlike the approach of Refs.~\cite{Akcay_et_al2012,Nori_pvt,Ba.al.10} which also requires the spacetime components of the self-force. 
As we already explained above, the use of \verb|Mathematica| coupled with the Teukolsky-MST-CCK approach is the other major reason for our tremendous improvement in accuracy.

From the Laurent series for $z_{(1)}^{e^2}(v)$ we can also obtain a series expansion for $\bar{d}(v)$ near the ISCO,
\begin{equation}\label{eq:dnearISCO}
\bar{d}(v) = \sum_{i=-2}^\infty d_i (1-6v)^i.
\end{equation}
For the divergent terms we find  $d_{-2}\lesssim 10^{-11}$ and  $d_{-1}\lesssim10^{-9}$, confirming that $\bar{d}(v)$ is indeed a regular function at the ISCO, as expected. For the finite part of Eq.~\eqref{eq:dnearISCO} we find
\begin{equation}\label{eq:dBars_coefficents}
\begin{split}
 d_{0} &= +0.666488(2),\\
 d_{1} &= -2.474180(9),\\
 d_{2} &= +4.436(2),\\
 d_{3} &= -6.073(2),\quad\text{and}\\
 d_{4} &= +7.92(7).
\end{split}
\end{equation}

Fig.~\ref{fig:dbar_overview} shows our numerical results together with the near-ISCO series expansion of $\bar{d}(v)$ given by Eq.~\eqref{eq:dnearISCO}. We also plot \BDG's 6.5pN approximation and their Pad\'{e} fit which, as shown, matches our data better in the strong-field regime. 
As can be seen, our error bars are too small to be distinguished even in the inset.
This is expected since our largest relative error for $\bar{d}(v)$ is $\approx 5\times 10^{-4}$.%Fig.~\ref{fig:d_bar_AbsDiff_vs_PN} shows that the absolute error in $\bar{d}(v)$ is $ \ord(10^{-5})$ at $v=1/6$ whereas the magnitude of $\bar{d}(v)$ is $\ord(1)$. 
The figure also shows that our near-ISCO expansion matches our near-ISCO data very well testifying to the quality of our numerical results for $\bar{d}(v)$ even just outside the ISCO.
The full numerical data for $\bar{d}(v)$ is provided verbatim in Table \ref{table:Numerical_DataI} of App.~A.

\subsection{The potential \texorpdfstring{$q(v)$}{q}}\label{sec:q_computation}
\begin{figure}[tb]
	\includegraphics[width=0.99\linewidth]{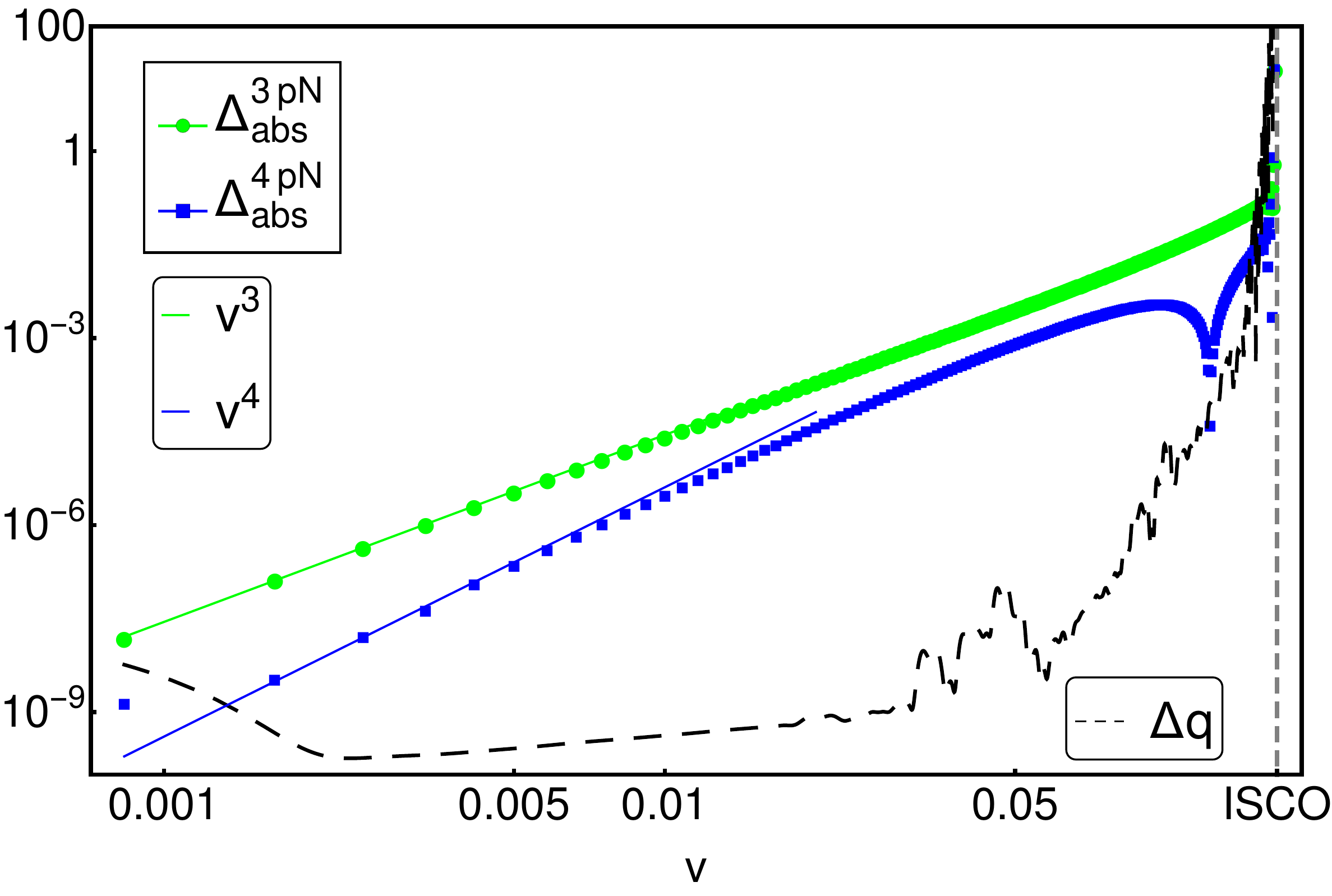}
	\caption{Absolute differences in $q(v)$ between our numerical data and Ref.~\cite{DJS2015}'s corresponding pN series. The green dots and blue squares represent the difference between our data and their 3, 4 pN expressions, respectively. The corresponding green and blue lines are fits that scale as $v^3, v^4$, respectively. 
	Note that according to Ref.~\cite{DJS2015}, the 3pN term scales as $v^2$ and the 4pN term as $v^3$ as such the absolute difference curves asymptotically scale as $v^3,v^4$. The black dashed curve ($\Delta q$) is our estimated error for $q(v)$.}% as $v\to 0$.}
\label{fig:q_AbsDiff_vs_PN}
\end{figure}
\begin{figure}[tb]
  \includegraphics[width=0.99\linewidth]{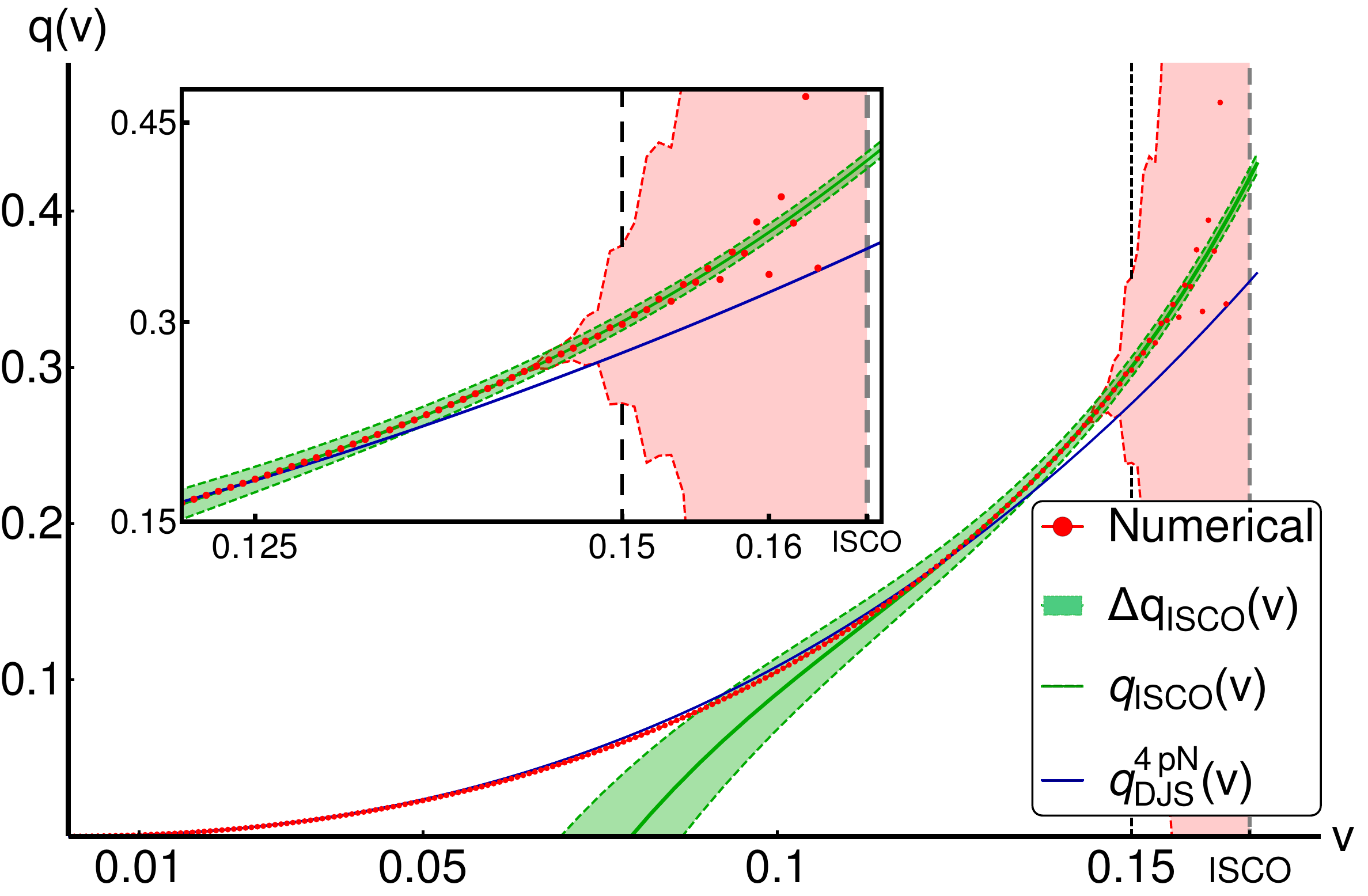}%
\caption{Numerical results for the potential $q(v)$ (red dots). The confidence region for these points is shaded in red. As $v$ approaches the ISCO our confidence sharply decreases due to large cancellations in Eq.~\eqref{eq:q_master}. 
In addition we show the known 4 pN expression for $q(v)$ (in blue) and the near-ISCO expansion from Eq.~\eqref{eq:qnearISCO} and its confidence interval (in green).
}
\label{fig:q_overview}
\end{figure}
Having explained in detail the computation of the various terms in Eq.~(\ref{eq:q_master}) we can now determine $q(v)$ across our range of $v$ values from $1/1200$ to $199/1200$. However, taking into account the  $(1-6v)^{-1,-3}$ behavior of $z_{(1)}^{e^2,e^4}(v)$ as $v\to 1/6$ and the explicit $(1-6v)^{-k},\: (k=1, \ldots, 5)$ coefficients, we see that many of the individual terms in  Eq.~(\ref{eq:q_master}) for $q(v)$ will diverge as $(1-6v)^{-5}$ as $v\to 1/6$. Nonetheless, it is well known that the EOB potentials $a(v),\bar{d}(v)$ and $q(v)$ are all regular at the ISCO \cite{Akcay_et_al2012} so this apparent divergence is an artifact of \AT's formulation. To test the behavior of $q(v)$  near the ISCO we write it as a Laurent series,
\begin{equation}\label{eq:qnearISCO}
	q(v) = \sum_{i=-5}^\infty q_i (1-6v)^i.
\end{equation}
By inserting the numerically obtained Laurent series for $z_{(1)}^{e^0}(v)$, $z_{(1)}^{e^2}(v)$, and $z_{(1)}^{e^4}(v)$ into Eq.~\eqref{eq:q_master} we obtain for the divergent terms
\begin{equation}
\begin{split}
 \abs{q_{-5}}	&\lesssim 10^{-11},\\
 \abs{q_{-4}}	&\lesssim 10^{-9},\\
 \abs{q_{-3}}	&\lesssim 10^{-7},\\
 \abs{q_{-2}}	&\lesssim 10^{-5}\text{,\qquad and}\\
 \abs{q_{-1}}	&\lesssim 10^{-3}.\\
\end{split}
\end{equation}
Thus, the $q$ potential indeed seems to be regular at the ISCO as expected. Assuming that the divergent part of the Laurent series vanishes identically we find for the regular part
\begin{equation}
\begin{split}
 q_{0} &= +0.421(9),\\
 q_{1} &= -1.447(9),\\
 q_{2} &= +2.62(9)\text{,\qquad and}\\
 q_{3} &= -2.6(3).
\end{split}
\end{equation}

With the regularity of $q(v)$ at least numerically established, we obtain it using Eq.~(\ref{eq:q_master}) rewritten in terms of the ISCO-regular functions $\tilde{z}_{(1)}^{e^2,\,e^4}(v)$, $\tilde{z}_{(1)}^{e^2}{}'(v)$, and $\tilde{z}_{(1)}^{e^2}{}''(v)$. 
In Fig.~\ref{fig:q_AbsDiff_vs_PN} we compare how well our numerical results match the 4pN expression of Ref.~\cite{DJS2015} by plotting the absolute difference $|q(v)-q^{4pN}_\mathsf{DJS}(v)|$. % Interestingly, we observe a sign change at $v\approx 0.12$. %Not that interesting residuals change sign quite commonly.
In the same figure, we also show our estimated numerical error for $q(v)$ which is $\lesssim 10^{-6}$ for $v\lesssim 1/10$. The apparent increase in our error at the $v\to 0$ end of our grid is due to the edge effects of finite differencing. That aside, our data is accurate enough to confidently detect %(at least visually) 
the expected asymptotic $v^4$ as $v\to 0$ behavior of the 4 pN residual, although this behavior has not settled enough to accurately determine the 5 pN coefficients. %{\bf Should we add the numerical values of the 5pN and 5pN log coefficients? I think this opens a new can of worms because we can now compare with the new DBG paper.}

In Fig.~\ref{fig:q_overview} we plot the full numerical results together with the 4 pN approximation from Ref.~\cite{DJS2015} and the near-ISCO expansion $q_\mathsf{ISCO}(v)$ from Eq.~(\ref{eq:qnearISCO}). 
Interestingly, all three almost coincide at $v\approx 0.12$, suggesting a good starting point for a simple analytic fit to the data (which we do not attempt here). 
As expected, the large numerical cancellations needed to remove the divergent $(1-6v)^{-k}, k=1,\ldots,5$ behavior of Eq.~\eqref{eq:q_master} at the ISCO cause a loss numerical precision in the strong field $v\gtrsim 0.15$. 
Nonetheless, in this regime, the near-ISCO expansion of Eq.~\eqref{eq:qnearISCO} provides results  with a $2\%$ accuracy as shown by the green confidence region of Fig.~\ref{fig:q_overview}. 
On the other hand, the confidence on our numerical value for $q(v)$ itself at the three-nearest-ISCO points degrades so significantly that the values have essentially no meaning. We nonetheless kept them in the presentation of our data to show our current limitations. The full numerical results for $q(v)$ can be found in Table \ref{table:Numerical_DataI}.

%\begin{figure}[tb]
%  \includegraphics[width=0.99\linewidth]{Figs/q_tilde_power_law_at_isco.pdf}%
%\caption{The power-law behavior of $\tilde{q}(v)\equiv (1-6v)^5 q(v)$ near the ISCO ($v=1/6$). The black line is our estimation for the leading-order behavior of $\tilde{q}(v)$ which implies that $q(v)$ is regular at the ISCO. We hit numerical noise for $v \gtrsim 0.16$ i.e. $\mathsf{y}_5 \lesssim 10^{-7}$.}
%\label{fig:q_tilde_power_law_at_isco}
%\end{figure}

\section{Discussion and conclusions}\label{sec:conclusions}
In this paper we have provided the first numerical calculation of the linear-in-mass-ratio EOB potential $q(v)$ in the range $0\leq v\leq 1/6$, using data from numerical self-force calculations. 
The key ingredient for this calculation is a relation between the so-called redshift invariant on slightly eccentric orbits and the various EOB potentials for compact (non-spinning) eccentric binaries, derived by Le Tiec in Ref.~\cite{LeTiec2015} using the eccentric generalization of the first law of binary mechanics. 
Our results for $q(v)$ are accurate to four to seven digits for most orbital separations, except in the region near the ISCO at $v=1/6$ where large numerical cancellations lead to a significant loss of precision by a few orders of magnitude. 
However, in that region we are able to extract the near-ISCO behavior of $q(v)$ as a Taylor series around $v=1/6$. 
At the same time we greatly improve on previous numerical determinations of $\bar{d}(v)$ in \cite{Akcay_et_al2012,Ba.al.12}: our strong-field results have twice as many significant digits.

One of the main hindrances in improving the numerical accuracy of $q(v)$ and $\bar{d}(v)$ is the singular nature of Eqs.~\eqref{eq:dBar_master} and \eqref{eq:q_master}, leading to large cancellations near the ISCO. 
As discussed in \BDG~this is related the loss of stable perturbed circular orbits below the ISCO, and is an inherent shortcoming of using the $e^2$ expansion of the redshift for determining the EOB potentials. 
As such this method could never probe the extremely strong-field regime of $1/6 \le v < 1/3$. This would require a very different approach based on extracting gauge-invariant information from hyperbolic orbits as detailed by Ref.~\cite{Damour:2009sm}. 
Unfortunately, these orbits are currently out of reach of both frequency and time-domain self-force computations. However, a comparable-mass ratio calculation was recently carried out successfully using full numerical relativity \cite{Damour:2014afa}.

Be that as it may, the near-ISCO expansions from our frequency-domain methods provide a first-ever partial look into the extremely strong field behavior of $q(v)$. Moreover, our values for $\bar{d}(v),q(v)$ are robust enough to provide accurate gravitational waveform templates.

The current results do not represent the limit of what is possible with our code accuracy-wise. In principle, the code used for calculating the GSF correction to the redshift can produce results at any desired accuracy, albeit at the cost of computation time. The main limiting factor is in the number of $l$ modes calculated. The results here are for a maximum of 40 $l$ modes. 
Each additional digit of accuracy in the strong-field regime would require about five additional $l$ modes, while computation times scale with at least $l^2$, possibly faster.

Currently, the most constraining factor is the accuracy of the finite difference derivatives used. These could simply be improved by producing a denser sampling in $v$, especially in the very strong-field region ($v\gtrsim 1/7$) where our accuracy is limited. Further improvements could be made by using pseudospectral methods on an adapted grid. 
Near the ISCO, we might ameliorate our current results by improving the near-ISCO expansion of the redshift functions. Currently, this expansion is what yields the most accurate results for $q(v)$ near the ISCO. This expansion could be improved significantly by calculating more dedicated data points very close to the ISCO.

Finally, for the derivatives of $z_{(1)}^{e^0}(v)$ in the strong-field regime $v \gtrsim 1/10$ we have relied on Padé fits to highly accurate circular-orbit data. If more dense strong-field data were available these fits could be improved significantly. For a dense enough grid, the desired accuracy could even be reached using finite difference derivatives.

\begin{acknowledgments}
SA thanks Alexandre Le Tiec, Niels Warburton, Barry Wardell, Chris Kavanagh, Nori Sago and Leor Barack. SA also gratefully acknowledges support from the Irish Research Council, funded under the National Development Plan for Ireland. 
MvdM was supported by the European Research Council under the European Union's Seventh Framework Programme (FP7/2007-2013)/ERC grant agreement no. 304978. The numerical results in this paper were obtained using the IRIDIS High Performance Computing Facility at 
the University of Southampton.
\end{acknowledgments}
\appendix
\section{Relations between Laurent coefficients of \texorpdfstring{$z_{(1)}^{e^0}$}{z1e0}, \texorpdfstring{$z_{(1)}^{e^2}$}{z1e2}, and \texorpdfstring{$z_{(1)}^{e^4}$}{z1e4}}
The fact that the potentials $\bar{d}(v)$ and $q(v)$ are expected to be regular functions while the expressions for them in Eqs.~\eqref{eq:dBar_master} and \eqref{eq:q_master} appear to be singular at the ISCO implies that relations must exist between the Laurent expansions of 
$z_{(1)}^{e^0}(v)$, $z_{(1)}^{e^2}(v)$, and $z_{(1)}^{e^4}(v)$. If in addition to Eqs.~\eqref{eq:ze2nearISCO} and \eqref{eq:ze4nearISCO} we write
\begin{equation}\label{eq:ze0nearISCO}
 z_{(1)}^{e^0}(v) = \sum_{i=0}^\infty c_i^{e^0} (1-6v)^i,
\end{equation}
then regularity of $\bar{d}(v)$ at the ISCO implies
\begin{subequations}
\begin{align}
 c_{-1}^{e^2} &=-\frac{1}{12}  c_{1}^{e^0},\quad\text{and}\\
 c_{0}^{e^2} &=\frac{11}{12}c_{1}^{e^0}+ \frac{1}{3}c_{2}^{e^0}.
\end{align}
\end{subequations}
Moreover, imposing regularity of $q(v)$ in Eq.\eqref{eq:q_master} yields
\begin{subequations}
\begin{align}
 c_{-3}^{e^4} &=-\frac{1}{384}  c_{1}^{e^0},\\ 
 c_{-2}^{e^4} &= \frac{1}{128}  c_{1}^{e^0},\\
 c_{-1}^{e^4} &=-\frac{23}{384}  c_{1}^{e^0} - \frac{7}{432}  c_{2}^{e^0} - \frac{7}{288}  c_{3}^{e^0} -\frac{7}{288}  c_{1}^{e^2},\\
 c_{ 0}^{e^4} &=\frac{2641}{6912}  c_{1}^{e^0} + \frac{179}{864}  c_{2}^{e^0} + \frac{41}{144}  c_{3}^{e^0} + \frac{43}{432}  c_{4}^{e^0}\\
 &\quad +\frac{23}{72}  c_{1}^{e^2}+ \frac{5}{288}  c_{2}^{e^2},\quad\text{and} \\
 c_{ 1}^{e^4} &= \frac{3971}{13824}  c_{1}^{e^0} + \frac{13}{108}  c_{2}^{e^0} - \frac{113}{96}  c_{3}^{e^0} - \frac{167}{216}  c_{4}^{e^0}\\
 &\quad  - \frac{55}{432}  c_{5}^{e^0} +\frac{53}{72}  c_{1}^{e^2}+ \frac{10}{9}  c_{2}^{e^2} + \frac{89}{288}  c_{3}^{e^2}.
\end{align}
\end{subequations}
The expansion of $z_{(1)}^{e^0}$ in Eq.~\eqref{eq:ze0nearISCO} can be determined numerically to high accuracy by sampling circular orbits close to the ISCO. For the leading coefficients we find,

\begin{subequations}
\begin{align}
 c_{0}^{e^0} &= +0.1480137546476(8),\\
 c_{1}^{e^0} &= -0.163746546807(2),\\
 c_{2}^{e^0} &= +0.100101531603(3),\\
 c_{3}^{e^0} &= -0.1851821013(3), \\
 c_{4}^{e^0} &= +0.2043911129(7),\\
 c_{5}^{e^0} &= -0.21100109(7),\quad\text{and}\\
 c_{6}^{e^0} &= +0.21978853(4).
\end{align}
\end{subequations}

This allows us to determine the singular parts of $z_{(1)}^{e^2}$ and $z_{(1)}^{e^4}$ to great accuracy.
\newpage
\begin{widetext}
\section{Numerical data}\label{sec:App_A}
%\clearpage
\renewcommand{\thefootnote}{\alph{footnote}} %hack to get PRD compliant footnotes in table below.
\setlength{\LTcapwidth}{0.7\linewidth}
\squeezetable
\begin{longtable*}[H]{cd{7}d{19}d{10}d{16}d{10}}
176/1200 & 6.818182 & 4.244142(3)  \times 10^{-1}  &  3.108  \times 10^{-7}  &2.80(9)  \times 10^{-1}  &  9.216  \times 10^{-3}   \kill
 \caption{Numerical values for $\bar{d}(v)$ and $q(v)$ together with estimates for their absolute errors. In strong field the values are supplemented with (more accurate) estimates based on the near ISCO expansions in Eqs.~\eqref{eq:dnearISCO} and \eqref{eq:qnearISCO}. 
 We would be happy to share electronic versions of our data with interested parties.
 \label{table:Numerical_DataI}
 }\\
 \hline\hline
 \multicolumn{1}{c}{$v$} & \multicolumn{1}{c}{$p =1/v$} & \multicolumn{1}{c}{$\bar{d}(v)$} & \multicolumn{1}{c}{$\Delta \bar{d}(v)$}  & \multicolumn{1}{c}{$q(v)$}  & \multicolumn{1}{c}{$\Delta q(v)$} \T\B \\
 \hline
 \endfirsthead
 \caption[]{(continued)}\\
% \toprule
 \hline\hline
 \multicolumn{1}{c}{$v$} & \multicolumn{1}{c}{$p =1/v$} & \multicolumn{1}{c}{$\bar{d}(v)$} & \multicolumn{1}{c}{$\Delta \bar{d}(v)$} & \multicolumn{1}{c}{$q(v)$}  & \multicolumn{1}{c}{$\Delta q(v)$} \T\B \\
 \hline\vspace{-8pt}
 \endhead
 \botrule
 \endfoot
 \botrule
 \endlastfoot
200/1200\footnotetext[1]{Value obtained from near ISCO expansion.} & 6.000000 & 6.66488(2) \times 10^{-1}\footnotemark[1]  &  1.7  \times 10^{-6}  & 0.421(9) \times 10^{-1}\footnotemark[1] & 8.6 \times 10^{-3}\T \\ 
199/1200 & 6.030151 & 6.542(3) \times 10^{-1}  &  2.5 \times 10^{-4}  & -1.85\nsd \times 10^{3} & 5.8 \times 10^{5} \\
	&	& 6.54228(2)\times 10^{-1} \footnotemark[1] & 1.7 \times 10^{-6} &  4.14(9)\times 10^{-1} \footnotemark[1] & 8.6 \times 10^{-3}	\\ 
198/1200 & 6.060606 & 6.4218(5)  \times 10^{-1}  &  5.3 \times 10^{-5}  & 2.22\nsd \times 10^{1} & 7.8 \times 10^{3}  \\ 
	&	& 6.42185(2)\times 10^{-1} \footnotemark[1] & 1.7 \times 10^{-6} &  4.07(9)\times 10^{-1} \footnotemark[1] & 8.6 \times 10^{-3}	\\ 
197/1200 & 6.091371 & 6.3035(2)  \times 10^{-1}  &  2.1 \times 10^{-5}  & -7.76\nsd \times 10^{-2} & 4.1 \times 10^{2}  \\ 
	&	& 6.30354(2)\times 10^{-1} \footnotemark[1] & 1.7 \times 10^{-6} &  4.00(9)\times 10^{-1} \footnotemark[1] & 8.6 \times 10^{-3}	\\ 
196/1200 & 6.122449 & 6.1873(1)  \times 10^{-1}  &  1.1 \times 10^{-5}  & 4.26\nsd \times 10^{-1} & 2.8 \times 10^{1} \\ 
	&	& 6.18732(2)\times 10^{-1} \footnotemark[1] & 1.9 \times 10^{-6} &  3.94(9)\times 10^{-1} \footnotemark[1] & 8.6 \times 10^{-3}	\\ 
195/1200 & 6.153846 & 6.07311(7)  \times 10^{-1}  &  7.2 \times 10^{-6}  & 4.99\nsd \times 10^{-1} & 1.2 \times 10^{1} \\
	&	& 6.07315(2)\times 10^{-1} \footnotemark[1] & 2.1 \times 10^{-6} &  3.87(9)\times 10^{-1} \footnotemark[1] & 8.6 \times 10^{-3}	\\ 
194/1200 & 6.185567 & 5.96095(5)  \times 10^{-1}  &  5.1 \times 10^{-6}  & 3.87\nsd \times 10^{-1}& 9.8 \times 10^{0} \\ 
	&	& 5.96098(3)\times 10^{-1} \footnotemark[1] & 2.6 \times 10^{-6} &  3.80(9)\times 10^{-1} \footnotemark[1] & 8.6 \times 10^{-3}	\\ 
193/1200 & 6.217617 & 5.85075(4)  \times 10^{-1}  &  3.9 \times 10^{-6}  & 4.00\nsd \times 10^{-1}& 3.6 \times 10^{0} \\ 
	&	& 5.85078(3)\times 10^{-1} \footnotemark[1] & 3.1 \times 10^{-6} &  3.74(9)\times 10^{-1} \footnotemark[1] & 8.6 \times 10^{-3}	\\ 
192/1200 & 6.250000 & 5.74248(3)  \times 10^{-1}  &  3.1 \times 10^{-6}  & 3.38\nsd \times 10^{-1}& 1.9 \times 10^{0} \\ 
	&	& 5.74250(4)\times 10^{-1} \footnotemark[1] & 3.8 \times 10^{-6} &  3.68(9)\times 10^{-1} \footnotemark[1] & 8.6 \times 10^{-3}	\\ 
191/1200 & 6.282723 & 5.63610(3)  \times 10^{-1}  &  2.6 \times 10^{-6}  & 3.77\nsd \times 10^{-1}& 1.4 \times 10^{0} \\ 
	&	& 5.63611(5)\times 10^{-1} \footnotemark[1] & 4.7 \times 10^{-6} &  3.61(9)\times 10^{-1} \footnotemark[1] & 8.6 \times 10^{-3}	\\ 
190/1200 & 6.315789 & 5.53157(2)  \times 10^{-1}  &  2.1 \times 10^{-6}  & 3.52\nsd \times 10^{-1}& 1.2 \times 10^{0} \\ 
	&	& 5.53158(6)\times 10^{-1} \footnotemark[1] & 5.6 \times 10^{-6} &  3.55(9)\times 10^{-1} \footnotemark[1] & 8.6 \times 10^{-3}	\\ 
189/1200 & 6.349206 & 5.42886(2)  \times 10^{-1}  &  1.8 \times 10^{-6}  & 3.53\nsd \times 10^{-1}& 5.8 \times 10^{-1} \\ 	
&	& 5.42886(7)\times 10^{-1} \footnotemark[1] & 6.8 \times 10^{-6} &  3.49(9)\times 10^{-1} \footnotemark[1] & 8.6 \times 10^{-3}	\\ 
188/1200 & 6.382979 & 5.32792(2)  \times 10^{-1}  &  1.5 \times 10^{-6}  & 3.32\nsd \times 10^{-1}& 3.8 \times 10^{-1} \\ 
&	& 5.32793(8)\times 10^{-1} \footnotemark[1] & 8.1 \times 10^{-6} &  3.43(9)\times 10^{-1} \footnotemark[1] & 8.6 \times 10^{-3}	\\ 
187/1200 & 6.417112 & 5.22874(1)  \times 10^{-1}  &  1.3 \times 10^{-6}  & 3.40\nsd \times 10^{-1}& 3.5 \times 10^{-1} \\ 
&	& 5.22874(10)\times 10^{-1} \footnotemark[1] & 9.6 \times 10^{-6} &  3.38(9)\times 10^{-1} \footnotemark[1] & 8.6 \times 10^{-3}	\\ 
186/1200 & 6.451613 & 5.13127(1)  \times 10^{-1}  &  1.1 \times 10^{-6}  & 3.30\nsd \times 10^{-1}& 2.9 \times 10^{-1} \\
&	& 5.1313(1)\times 10^{-1} \footnotemark[1] & 1.1 \times 10^{-5} &  3.32(9)\times 10^{-1} \footnotemark[1] & 8.6 \times 10^{-3}	\\
185/1200 & 6.486486 & 5.03549(1)  \times 10^{-1}  &  1.0 \times 10^{-6}  & 3.28\nsd \times 10^{-1}& 1.6  \times 10^{-1}\\ 
&	& 5.0355(1)\times 10^{-1} \footnotemark[1] & 1.3 \times 10^{-5} &  3.27(9)\times 10^{-1} \footnotemark[1] & 8.6 \times 10^{-3}	\\
184/1200 & 6.521739 & 4.941351(9)  \times 10^{-1}  &  8.8  \times 10^{-7}  & 3.16\nsd \times 10^{-1}& 1.2  \times 10^{-1}\\ 
&	& 4.9413(2)\times 10^{-1} \footnotemark[1] & 1.6 \times 10^{-5} &  3.21(9)\times 10^{-1} \footnotemark[1] & 8.6 \times 10^{-3}	\\
183/1200 & 6.557377 & 4.848836(8)  \times 10^{-1}  &  7.6  \times 10^{-7}  & 3.17\nsd \times 10^{-1} & 1.2 \times 10^{-1} \\ 
&	& 4.8488(2)\times 10^{-1} \footnotemark[1] & 1.9 \times 10^{-5} &  3.16(9)\times 10^{-1} \footnotemark[1] & 8.6 \times 10^{-3}	\\
182/1200 & 6.593407 & 4.757911(7)  \times 10^{-1}  &  6.7  \times 10^{-7}  & 3.09\nsd \times 10^{-1}& 1.2 \times 10^{-1} \\ 
&	& 4.7579(2)\times 10^{-1} \footnotemark[1] & 2.3 \times 10^{-5} &  3.10(9)\times 10^{-1} \footnotemark[1] & 8.7 \times 10^{-3}	\\
181/1200 & 6.629834 & 4.668546(6)  \times 10^{-1}  &  5.8  \times 10^{-7}  & 3.06(69) \times 10^{-1}& 6.9 \times 10^{-2} \\ 
180/1200 & 6.666667 & 4.580713(5)  \times 10^{-1}  &  5.1  \times 10^{-7}  & 2.98(59) \times 10^{-1}& 5.9 \times 10^{-2} \\ 
179/1200 & 6.703911 & 4.494383(5)  \times 10^{-1}  &  4.5  \times 10^{-7}  & 2.96(58) \times 10^{-1}& 5.8 \times 10^{-2} \\ 
178/1200 & 6.741573 & 4.409529(4)  \times 10^{-1}  &  4.0  \times 10^{-7}  & 2.89(20) \times 10^{-1}& 2.0 \times 10^{-2} \\ 
177/1200 & 6.779661 & 4.326124(4)  \times 10^{-1}  &  3.5  \times 10^{-7}  & 2.86(18) \times 10^{-1}& 1.8 \times 10^{-2} \\ 
176/1200 & 6.818182 & 4.244142(3)  \times 10^{-1}  &  3.1  \times 10^{-7}  &2.80(9)  \times 10^{-1}  &  9.2  \times 10^{-3}  \\ 
175/1200 & 6.857143 & 4.163557(3)  \times 10^{-1}  &  2.7  \times 10^{-7}  &2.76(7)  \times 10^{-1}  &  6.7  \times 10^{-3}  \\ 
174/1200 & 6.896552 & 4.084344(2)  \times 10^{-1}  &  2.4  \times 10^{-7}  &2.72(7)  \times 10^{-1}  &  6.6  \times 10^{-3}  \\ 
173/1200 & 6.936416 & 4.006479(2)  \times 10^{-1}  &  2.1  \times 10^{-7}  &2.67(2)  \times 10^{-1}  &  1.9  \times 10^{-3}  \\ 
172/1200 & 6.976744 & 3.929937(2)  \times 10^{-1}  &  1.9  \times 10^{-7}  &2.630(9)  \times 10^{-1}  &  9.5  \times 10^{-4}  \\ 
171/1200 & 7.017544 & 3.854696(2)  \times 10^{-1}  &  1.7  \times 10^{-7}  &2.583(9)  \times 10^{-1}  &  9.4  \times 10^{-4}  \\ 
170/1200 & 7.058824 & 3.780731(1)  \times 10^{-1}  &  1.5  \times 10^{-7}  &2.54(1)  \times 10^{-1}  &  1.0  \times 10^{-3}  \\ 
169/1200 & 7.100592 & 3.708022(1)  \times 10^{-1}  &  1.3  \times 10^{-7}  &2.500(7)  \times 10^{-1}  &  7.0  \times 10^{-4}  \\ 
168/1200 & 7.142857 & 3.636545(1)  \times 10^{-1}  &  1.2  \times 10^{-7}  &2.462(6)  \times 10^{-1}  &  6.1  \times 10^{-4}  \\ 
167/1200 & 7.185629 & 3.566279(1)  \times 10^{-1}  &  1.0  \times 10^{-7}  &2.419(6)  \times 10^{-1}  &  6.5  \times 10^{-4}  \\ 
166/1200 & 7.228916 & 3.4972032(9)  \times 10^{-1}  &  9.2  \times 10^{-8}  &2.380(3)  \times 10^{-1}  &  3.0  \times 10^{-4}  \\ 
165/1200 & 7.272727 & 3.4292972(8)  \times 10^{-1}  &  8.2  \times 10^{-8}  &2.340(2)  \times 10^{-1}  &  2.1  \times 10^{-4}  \\ 
164/1200 & 7.317073 & 3.3625406(7)  \times 10^{-1}  &  7.2  \times 10^{-8}  &2.304(2)  \times 10^{-1}  &  1.9  \times 10^{-4}  \\ 
163/1200 & 7.361963 & 3.2969137(6)  \times 10^{-1}  &  6.4  \times 10^{-8}  &2.264(3)  \times 10^{-1}  &  2.6  \times 10^{-4}  \\ 
162/1200 & 7.407407 & 3.2323971(6)  \times 10^{-1}  &  5.7  \times 10^{-8}  &2.227(6)  \times 10^{-1}  &  6.4  \times 10^{-4}  \\ 
161/1200 & 7.453416 & 3.1689719(5)  \times 10^{-1}  &  5.1  \times 10^{-8}  &2.190(4)  \times 10^{-1}  &  4.5  \times 10^{-4}  \\ 
160/1200 & 7.500000 & 3.1066197(5)  \times 10^{-1}  &  4.5  \times 10^{-8}  &2.155(4)  \times 10^{-1}  &  4.0  \times 10^{-4}  \\ 
159/1200 & 7.547170 & 3.0453222(4)  \times 10^{-1}  &  4.0  \times 10^{-8}  &2.118(5)  \times 10^{-1}  &  4.7  \times 10^{-4}  \\ 
158/1200 & 7.594937 & 2.9850617(4)  \times 10^{-1}  &  3.6  \times 10^{-8}  &2.083(4)  \times 10^{-1}  &  3.7  \times 10^{-4}  \\ 
157/1200 & 7.643312 & 2.9258207(3)  \times 10^{-1}  &  3.2  \times 10^{-8}  &2.048(3)  \times 10^{-1}  &  2.6  \times 10^{-4}  \\ 
156/1200 & 7.692308 & 2.8675821(3)  \times 10^{-1}  &  2.8  \times 10^{-8}  &2.015(2)  \times 10^{-1}  &  2.3  \times 10^{-4}  \\ 
155/1200 & 7.741935 & 2.8103293(2)  \times 10^{-1}  &  2.5  \times 10^{-8}  &1.981(3)  \times 10^{-1}  &  2.7  \times 10^{-4}  \\ 
154/1200 & 7.792208 & 2.7540459(2)  \times 10^{-1}  &  2.2  \times 10^{-8}  &1.948(1)  \times 10^{-1}  &  1.5  \times 10^{-4}  \\ 
153/1200 & 7.843137 & 2.6987158(2)  \times 10^{-1}  &  2.0  \times 10^{-8}  &1.915(1)  \times 10^{-1}  &  9.9  \times 10^{-5}  \\ 
152/1200 & 7.894737 & 2.6443231(2)  \times 10^{-1}  &  1.8  \times 10^{-8}  &1.8829(8)  \times 10^{-1}  &  8.4  \times 10^{-5}  \\ 
151/1200 & 7.947020 & 2.5908526(2)  \times 10^{-1}  &  1.6  \times 10^{-8}  &1.851(1)  \times 10^{-1}  &  10.0  \times 10^{-5}  \\ 
150/1200 & 8.000000 & 2.5382891(1)  \times 10^{-1}  &  1.4  \times 10^{-8}  &1.820(1)  \times 10^{-1}  &  1.1  \times 10^{-4}  \\ 
149/1200 & 8.053691 & 2.4866177(1)  \times 10^{-1}  &  1.2  \times 10^{-8}  &1.788(1)  \times 10^{-1}  &  1.0  \times 10^{-4}  \\ 
148/1200 & 8.108108 & 2.4358239(1)  \times 10^{-1}  &  1.1  \times 10^{-8}  &1.7583(9)  \times 10^{-1}  &  8.6  \times 10^{-5}  \\ 
147/1200 & 8.163265 & 2.3858934(1)  \times 10^{-1}  &  9.8  \times 10^{-9}  &1.7279(9)  \times 10^{-1}  &  9.0  \times 10^{-5}  \\ 
146/1200 & 8.219178 & 2.33681219(9)  \times 10^{-1}  &  8.7  \times 10^{-9}  &1.6984(7)  \times 10^{-1}  &  7.3  \times 10^{-5}  \\ 
145/1200 & 8.275862 & 2.28856659(8)  \times 10^{-1}  &  7.8  \times 10^{-9}  &1.6691(4)  \times 10^{-1}  &  3.8  \times 10^{-5}  \\ 
144/1200 & 8.333333 & 2.24114312(7)  \times 10^{-1}  &  6.9  \times 10^{-9}  &1.6404(3)  \times 10^{-1}  &  3.3  \times 10^{-5}  \\ 
143/1200 & 8.391608 & 2.19452857(6)  \times 10^{-1}  &  6.2  \times 10^{-9}  &1.6118(4)  \times 10^{-1}  &  3.8  \times 10^{-5}  \\ 
142/1200 & 8.450704 & 2.14870997(5)  \times 10^{-1}  &  5.5  \times 10^{-9}  &1.5838(5)  \times 10^{-1}  &  4.7  \times 10^{-5}  \\ 
141/1200 & 8.510638 & 2.10367462(5)  \times 10^{-1}  &  4.9  \times 10^{-9}  &1.5561(3)  \times 10^{-1}  &  3.4  \times 10^{-5}  \\ 
140/1200 & 8.571429 & 2.05941003(4)  \times 10^{-1}  &  4.4  \times 10^{-9}  &1.5288(3)  \times 10^{-1}  &  3.0  \times 10^{-5}  \\ 
139/1200 & 8.633094 & 2.01590395(4)  \times 10^{-1}  &  3.9  \times 10^{-9}  &1.5019(4)  \times 10^{-1}  &  3.6  \times 10^{-5}  \\ 
138/1200 & 8.695652 & 1.97314437(4)  \times 10^{-1}  &  3.5  \times 10^{-9}  &1.4754(2)  \times 10^{-1}  &  2.1  \times 10^{-5}  \\ 
137/1200 & 8.759124 & 1.93111949(3)  \times 10^{-1}  &  3.1  \times 10^{-9}  &1.4492(1)  \times 10^{-1}  &  1.5  \times 10^{-5}  \\ 
136/1200 & 8.823529 & 1.88981773(3)  \times 10^{-1}  &  2.8  \times 10^{-9}  &1.4233(1)  \times 10^{-1}  &  1.4  \times 10^{-5}  \\ 
135/1200 & 8.888889 & 1.84922771(3)  \times 10^{-1}  &  2.5  \times 10^{-9}  &1.3979(2)  \times 10^{-1}  &  1.7  \times 10^{-5}  \\ 
134/1200 & 8.955224 & 1.80933828(2)  \times 10^{-1}  &  2.3  \times 10^{-9}  &1.3727(2)  \times 10^{-1}  &  1.8  \times 10^{-5}  \\ 
133/1200 & 9.022556 & 1.77013848(2)  \times 10^{-1}  &  2.0  \times 10^{-9}  &1.3479(1)  \times 10^{-1}  &  1.3  \times 10^{-5}  \\ 
132/1200 & 9.090909 & 1.73161755(2)  \times 10^{-1}  &  1.9  \times 10^{-9}  &1.3234(1)  \times 10^{-1}  &  1.2  \times 10^{-5}  \\ 
131/1200 & 9.160305 & 1.69376491(2)  \times 10^{-1}  &  1.8  \times 10^{-9}  &1.2993(2)  \times 10^{-1}  &  1.5  \times 10^{-5}  \\ 
130/1200 & 9.230769 & 1.65657019(2)  \times 10^{-1}  &  1.6  \times 10^{-9}  &1.2755(2)  \times 10^{-1}  &  1.5  \times 10^{-5}  \\ 
129/1200 & 9.302326 & 1.62002320(2)  \times 10^{-1}  &  1.5  \times 10^{-9}  &1.2520(1)  \times 10^{-1}  &  1.1  \times 10^{-5}  \\ 
128/1200 & 9.375000 & 1.58411391(1)  \times 10^{-1}  &  1.4  \times 10^{-9}  &1.2289(1)  \times 10^{-1}  &  1.0  \times 10^{-5}  \\ 
127/1200 & 9.448819 & 1.54883250(1)  \times 10^{-1}  &  1.3  \times 10^{-9}  &1.2060(1)  \times 10^{-1}  &  1.3  \times 10^{-5}  \\ 
126/1200 & 9.523810 & 1.51416930(1)  \times 10^{-1}  &  1.2  \times 10^{-9}  &1.18345(7)  \times 10^{-1}  &  7.1  \times 10^{-6}  \\ 
125/1200 & 9.600000 & 1.48011481(1)  \times 10^{-1}  &  1.1  \times 10^{-9}  &1.16120(5)  \times 10^{-1}  &  4.8  \times 10^{-6}  \\ 
124/1200 & 9.677419 & 1.44665971(1)  \times 10^{-1}  &  1.0  \times 10^{-9}  &1.13927(5)  \times 10^{-1}  &  4.7  \times 10^{-6}  \\ 
123/1200 & 9.756098 & 1.41379484(1)  \times 10^{-1}  &  9.6  \times 10^{-10}  &1.11761(8)  \times 10^{-1}  &  8.3  \times 10^{-6}  \\ 
122/1200 & 9.836066 & 1.381511178(9)  \times 10^{-1}  &  8.9  \times 10^{-10}  &1.0963(2)  \times 10^{-1}  &  2.2  \times 10^{-5}  \\ 
121/1200 & 9.917355 & 1.349799879(8)  \times 10^{-1}  &  8.2  \times 10^{-10}  &1.0752(2)  \times 10^{-1}  &  1.6  \times 10^{-5}  \\ 
120/1200 & 10.000000 & 1.318652241(8)  \times 10^{-1}  &  7.7  \times 10^{-10}  &1.0544(2)  \times 10^{-1}  &  1.6  \times 10^{-5}  \\ 
119/1200 & 10.084034 & 1.288059712(7)  \times 10^{-1}  &  6.9  \times 10^{-10}  &1.0339(2)  \times 10^{-1}  &  1.9  \times 10^{-5}  \\ 
118/1200 & 10.169492 & 1.258013887(6)  \times 10^{-1}  &  6.4  \times 10^{-10}  &1.01369(8)  \times 10^{-1}  &  8.1  \times 10^{-6}  \\ 
117/1200 & 10.256410 & 1.228506502(6)  \times 10^{-1}  &  5.9  \times 10^{-10}  &9.9374(5)  \times 10^{-2}  &  4.9  \times 10^{-6}  \\ 
116/1200 & 10.344828 & 1.199529432(5)  \times 10^{-1}  &  5.4  \times 10^{-10}  &9.7408(5)  \times 10^{-2}  &  4.6  \times 10^{-6}  \\ 
115/1200 & 10.434783 & 1.171074691(5)  \times 10^{-1}  &  4.9  \times 10^{-10}  &9.5466(6)  \times 10^{-2}  &  5.5  \times 10^{-6}  \\ 
114/1200 & 10.526316 & 1.143134425(4)  \times 10^{-1}  &  4.5  \times 10^{-10}  &9.3552(2)  \times 10^{-2}  &  1.7  \times 10^{-6}  \\ 
113/1200 & 10.619469 & 1.115700911(4)  \times 10^{-1}  &  4.1  \times 10^{-10}  &9.16644(9)  \times 10^{-2}  &  8.6  \times 10^{-7}  \\ 
112/1200 & 10.714286 & 1.088766555(4)  \times 10^{-1}  &  3.7  \times 10^{-10}  &8.98010(7)  \times 10^{-2}  &  7.5  \times 10^{-7}  \\ 
111/1200 & 10.810811 & 1.062323890(3)  \times 10^{-1}  &  3.4  \times 10^{-10}  &8.7966(1)  \times 10^{-2}  &  1.1  \times 10^{-6}  \\ 
110/1200 & 10.909091 & 1.036365569(3)  \times 10^{-1}  &  3.1  \times 10^{-10}  &8.6154(2)  \times 10^{-2}  &  2.4  \times 10^{-6}  \\ 
109/1200 & 11.009174 & 1.010884369(3)  \times 10^{-1}  &  2.8  \times 10^{-10}  &8.4368(2)  \times 10^{-2}  &  1.8  \times 10^{-6}  \\ 
108/1200 & 11.111111 & 9.85873182(3)  \times 10^{-2}  &  2.7  \times 10^{-10}  &8.2606(2)  \times 10^{-2}  &  1.7  \times 10^{-6}  \\ 
107/1200 & 11.214953 & 9.61325020(3)  \times 10^{-2}  &  2.6  \times 10^{-10}  &8.0870(2)  \times 10^{-2}  &  2.2  \times 10^{-6}  \\ 
106/1200 & 11.320755 & 9.37233004(3)  \times 10^{-2}  &  2.5  \times 10^{-10}  &7.9157(2)  \times 10^{-2}  &  1.6  \times 10^{-6}  \\ 
105/1200 & 11.428571 & 9.13590369(2)  \times 10^{-2}  &  2.4  \times 10^{-10}  &7.7468(1)  \times 10^{-2}  &  1.2  \times 10^{-6}  \\ 
104/1200 & 11.538462 & 8.90390459(2)  \times 10^{-2}  &  2.3  \times 10^{-10}  &7.5803(1)  \times 10^{-2}  &  1.1  \times 10^{-6}  \\ 
103/1200 & 11.650485 & 8.67626722(2)  \times 10^{-2}  &  2.3  \times 10^{-10}  &7.4161(1)  \times 10^{-2}  &  1.4  \times 10^{-6}  \\ 
102/1200 & 11.764706 & 8.45292715(2)  \times 10^{-2}  &  2.2  \times 10^{-10}  &7.25426(6)  \times 10^{-2}  &  6.2  \times 10^{-7}  \\ 
101/1200 & 11.881188 & 8.23382094(2)  \times 10^{-2}  &  2.1  \times 10^{-10}  &7.09474(4)  \times 10^{-2}  &  4.0  \times 10^{-7}  \\ 
100/1200 & 12.000000 & 8.01888618(2)  \times 10^{-2}  &  2.0  \times 10^{-10}  &6.93734(4)  \times 10^{-2}  &  3.8  \times 10^{-7}  \\ 
99/1200 & 12.121212 & 7.80806143(2)  \times 10^{-2}  &  2.0  \times 10^{-10}  &6.78241(5)  \times 10^{-2}  &  4.6  \times 10^{-7}  \\ 
98/1200 & 12.244898 & 7.60128623(2)  \times 10^{-2}  &  1.9  \times 10^{-10}  &6.62954(2)  \times 10^{-2}  &  2.4  \times 10^{-7}  \\ 
97/1200 & 12.371134 & 7.39850107(2)  \times 10^{-2}  &  1.8  \times 10^{-10}  &6.47895(2)  \times 10^{-2}  &  1.7  \times 10^{-7}  \\ 
96/1200 & 12.500000 & 7.19964734(2)  \times 10^{-2}  &  1.8  \times 10^{-10}  &6.33046(2)  \times 10^{-2}  &  1.5  \times 10^{-7}  \\ 
95/1200 & 12.631579 & 7.00466739(2)  \times 10^{-2}  &  1.7  \times 10^{-10}  &6.18422(2)  \times 10^{-2}  &  1.7  \times 10^{-7}  \\ 
94/1200 & 12.765957 & 6.81350442(2)  \times 10^{-2}  &  1.7  \times 10^{-10}  &6.04006(1)  \times 10^{-2}  &  1.1  \times 10^{-7}  \\ 
93/1200 & 12.903226 & 6.62610253(2)  \times 10^{-2}  &  1.6  \times 10^{-10}  &5.898042(8)  \times 10^{-2}  &  7.6  \times 10^{-8}  \\ 
92/1200 & 13.043478 & 6.44240668(2)  \times 10^{-2}  &  1.6  \times 10^{-10}  &5.758081(8)  \times 10^{-2}  &  8.3  \times 10^{-8}  \\ 
91/1200 & 13.186813 & 6.26236266(2)  \times 10^{-2}  &  1.5  \times 10^{-10}  &5.62025(1)  \times 10^{-2}  &  9.9  \times 10^{-8}  \\ 
90/1200 & 13.333333 & 6.08591710(2)  \times 10^{-2}  &  1.5  \times 10^{-10}  &5.484439(8)  \times 10^{-2}  &  8.0  \times 10^{-8}  \\ 
89/1200 & 13.483146 & 5.91301745(1)  \times 10^{-2}  &  1.5  \times 10^{-10}  &5.350685(7)  \times 10^{-2}  &  7.0  \times 10^{-8}  \\ 
88/1200 & 13.636364 & 5.74361193(1)  \times 10^{-2}  &  1.4  \times 10^{-10}  &5.218922(7)  \times 10^{-2}  &  7.0  \times 10^{-8}  \\ 
87/1200 & 13.793103 & 5.57764956(1)  \times 10^{-2}  &  1.4  \times 10^{-10}  &5.089195(7)  \times 10^{-2}  &  7.3  \times 10^{-8}  \\ 
86/1200 & 13.953488 & 5.41508012(1)  \times 10^{-2}  &  1.4  \times 10^{-10}  &4.961429(5)  \times 10^{-2}  &  5.2  \times 10^{-8}  \\ 
85/1200 & 14.117647 & 5.25585414(1)  \times 10^{-2}  &  1.3  \times 10^{-10}  &4.835633(5)  \times 10^{-2}  &  5.1  \times 10^{-8}  \\ 
84/1200 & 14.285714 & 5.09992290(1)  \times 10^{-2}  &  1.3  \times 10^{-10}  &4.711760(5)  \times 10^{-2}  &  4.8  \times 10^{-8}  \\ 
83/1200 & 14.457831 & 4.94723840(1)  \times 10^{-2}  &  1.3  \times 10^{-10}  &4.589865(4)  \times 10^{-2}  &  4.3  \times 10^{-8}  \\ 
82/1200 & 14.634146 & 4.79775334(1)  \times 10^{-2}  &  1.2  \times 10^{-10}  &4.469850(4)  \times 10^{-2}  &  4.1  \times 10^{-8}  \\ 
81/1200 & 14.814815 & 4.65142112(1)  \times 10^{-2}  &  1.2  \times 10^{-10}  &4.351744(4)  \times 10^{-2}  &  3.6  \times 10^{-8}  \\ 
80/1200 & 15.000000 & 4.50819583(1)  \times 10^{-2}  &  1.2  \times 10^{-10}  &4.235493(3)  \times 10^{-2}  &  3.0  \times 10^{-8}  \\ 
79/1200 & 15.189873 & 4.36803223(1)  \times 10^{-2}  &  1.1  \times 10^{-10}  &4.121148(3)  \times 10^{-2}  &  2.6  \times 10^{-8}  \\ 
78/1200 & 15.384615 & 4.23088575(1)  \times 10^{-2}  &  1.1  \times 10^{-10}  &4.008630(2)  \times 10^{-2}  &  1.6  \times 10^{-8}  \\ 
77/1200 & 15.584416 & 4.09671244(1)  \times 10^{-2}  &  1.1  \times 10^{-10}  &3.897949(1)  \times 10^{-2}  &  1.2  \times 10^{-8}  \\ 
76/1200 & 15.789474 & 3.96546901(1)  \times 10^{-2}  &  1.1  \times 10^{-10}  &3.789081(1)  \times 10^{-2}  &  1.2  \times 10^{-8}  \\ 
75/1200 & 16.000000 & 3.83711278(1)  \times 10^{-2}  &  1.0  \times 10^{-10}  &3.682027(2)  \times 10^{-2}  &  1.9  \times 10^{-8}  \\ 
74/1200 & 16.216216 & 3.71160169(1)  \times 10^{-2}  &  1.0  \times 10^{-10}  &3.576760(2)  \times 10^{-2}  &  1.5  \times 10^{-8}  \\ 
73/1200 & 16.438356 & 3.58889428(1)  \times 10^{-2}  &  10.0  \times 10^{-11}  &3.473259(1)  \times 10^{-2}  &  1.4  \times 10^{-8}  \\ 
72/1200 & 16.666667 & 3.46894968(1)  \times 10^{-2}  &  9.7  \times 10^{-11}  &3.371535(1)  \times 10^{-2}  &  1.5  \times 10^{-8}  \\ 
71/1200 & 16.901408 & 3.351727590(9)  \times 10^{-2}  &  9.5  \times 10^{-11}  &3.271546(1)  \times 10^{-2}  &  1.1  \times 10^{-8}  \\ 
70/1200 & 17.142857 & 3.237188287(9)  \times 10^{-2}  &  9.2  \times 10^{-11}  &3.1733009(5)  \times 10^{-2}  &  4.9  \times 10^{-9}  \\ 
69/1200 & 17.391304 & 3.125292606(9)  \times 10^{-2}  &  9.0  \times 10^{-11}  &3.0767676(4)  \times 10^{-2}  &  3.9  \times 10^{-9}  \\ 
68/1200 & 17.647059 & 3.016001929(9)  \times 10^{-2}  &  8.8  \times 10^{-11}  &2.9819610(5)  \times 10^{-2}  &  4.5  \times 10^{-9}  \\ 
67/1200 & 17.910448 & 2.909278175(9)  \times 10^{-2}  &  8.6  \times 10^{-11}  &2.8888264(4)  \times 10^{-2}  &  4.3  \times 10^{-9}  \\ 
66/1200 & 18.181818 & 2.805083796(8)  \times 10^{-2}  &  8.3  \times 10^{-11}  &2.797394(1)  \times 10^{-2}  &  10.0  \times 10^{-9}  \\ 
65/1200 & 18.461538 & 2.703381757(8)  \times 10^{-2}  &  8.1  \times 10^{-11}  &2.7076162(8)  \times 10^{-2}  &  8.3  \times 10^{-9}  \\ 
64/1200 & 18.750000 & 2.604135538(8)  \times 10^{-2}  &  7.9  \times 10^{-11}  &2.6195279(8)  \times 10^{-2}  &  8.2  \times 10^{-9}  \\ 
63/1200 & 19.047619 & 2.507309116(8)  \times 10^{-2}  &  7.7  \times 10^{-11}  &2.533046(2)  \times 10^{-2}  &  1.5  \times 10^{-8}  \\ 
62/1200 & 19.354839 & 2.412866962(7)  \times 10^{-2}  &  7.5  \times 10^{-11}  &2.448230(4)  \times 10^{-2}  &  4.3  \times 10^{-8}  \\ 
61/1200 & 19.672131 & 2.320774029(7)  \times 10^{-2}  &  7.3  \times 10^{-11}  &2.365015(3)  \times 10^{-2}  &  3.5  \times 10^{-8}  \\ 
60/1200 & 20.000000 & 2.230995743(7)  \times 10^{-2}  &  7.1  \times 10^{-11}  &2.283454(3)  \times 10^{-2}  &  3.4  \times 10^{-8}  \\ 
59/1200 & 20.338983 & 2.143497997(7)  \times 10^{-2}  &  6.9  \times 10^{-11}  &2.203441(5)  \times 10^{-2}  &  4.9  \times 10^{-8}  \\ 
58/1200 & 20.689655 & 2.058247141(7)  \times 10^{-2}  &  6.7  \times 10^{-11}  &2.12506(1)  \times 10^{-2}  &  1.0  \times 10^{-7}  \\ 
57/1200 & 21.052632 & 1.975209975(6)  \times 10^{-2}  &  6.5  \times 10^{-11}  &2.048224(8)  \times 10^{-2}  &  8.2  \times 10^{-8}  \\ 
56/1200 & 21.428571 & 1.894353741(6)  \times 10^{-2}  &  6.3  \times 10^{-11}  &1.973028(8)  \times 10^{-2}  &  8.0  \times 10^{-8}  \\ 
55/1200 & 21.818182 & 1.815646108(6)  \times 10^{-2}  &  6.1  \times 10^{-11}  &1.899298(9)  \times 10^{-2}  &  9.3  \times 10^{-8}  \\ 
54/1200 & 22.222222 & 1.739055180(6)  \times 10^{-2}  &  5.9  \times 10^{-11}  &1.827182(3)  \times 10^{-2}  &  3.2  \times 10^{-8}  \\ 
53/1200 & 22.641509 & 1.664549474(6)  \times 10^{-2}  &  5.7  \times 10^{-11}  &1.756586(2)  \times 10^{-2}  &  1.7  \times 10^{-8}  \\ 
52/1200 & 23.076923 & 1.592097917(6)  \times 10^{-2}  &  5.5  \times 10^{-11}  &1.687519(2)  \times 10^{-2}  &  1.6  \times 10^{-8}  \\ 
51/1200 & 23.529412 & 1.521669841(5)  \times 10^{-2}  &  5.3  \times 10^{-11}  &1.619978(2)  \times 10^{-2}  &  2.0  \times 10^{-8}  \\ 
50/1200 & 24.000000 & 1.453234973(5)  \times 10^{-2}  &  5.1  \times 10^{-11}  &1.553945(2)  \times 10^{-2}  &  1.9  \times 10^{-8}  \\ 
49/1200 & 24.489796 & 1.386763431(5)  \times 10^{-2}  &  5.0  \times 10^{-11}  &1.489413(2)  \times 10^{-2}  &  1.5  \times 10^{-8}  \\ 
48/1200 & 25.000000 & 1.322225714(5)  \times 10^{-2}  &  4.8  \times 10^{-11}  &1.426374(2)  \times 10^{-2}  &  1.5  \times 10^{-8}  \\ 
47/1200 & 25.531915 & 1.259592694(5)  \times 10^{-2}  &  4.6  \times 10^{-11}  &1.364819(2)  \times 10^{-2}  &  1.7  \times 10^{-8}  \\ 
46/1200 & 26.086957 & 1.198835617(4)  \times 10^{-2}  &  4.5  \times 10^{-11}  &1.3047396(5)  \times 10^{-2}  &  5.4  \times 10^{-9}  \\ 
45/1200 & 26.666667 & 1.139926086(4)  \times 10^{-2}  &  4.3  \times 10^{-11}  &1.2461250(3)  \times 10^{-2}  &  2.6  \times 10^{-9}  \\ 
44/1200 & 27.272727 & 1.082836061(4)  \times 10^{-2}  &  4.1  \times 10^{-11}  &1.1889634(2)  \times 10^{-2}  &  2.4  \times 10^{-9}  \\ 
43/1200 & 27.906977 & 1.027537853(4)  \times 10^{-2}  &  4.0  \times 10^{-11}  &1.1332589(4)  \times 10^{-2}  &  3.7  \times 10^{-9}  \\ 
42/1200 & 28.571429 & 9.74004114(4)  \times 10^{-3}  &  3.8  \times 10^{-11}  &1.078990(1)  \times 10^{-2}  &  9.6  \times 10^{-9}  \\ 
41/1200 & 29.268293 & 9.22207833(4)  \times 10^{-3}  &  3.7  \times 10^{-11}  &1.0261541(8)  \times 10^{-2}  &  8.3  \times 10^{-9}  \\ 
40/1200 & 30.000000 & 8.72122331(4)  \times 10^{-3}  &  3.5  \times 10^{-11}  &9.747377(8)  \times 10^{-3}  &  8.2  \times 10^{-9}  \\ 
39/1200 & 30.769231 & 8.23721251(3)  \times 10^{-3}  &  3.4  \times 10^{-11}  &9.247448(9)  \times 10^{-3}  &  9.0  \times 10^{-9}  \\ 
38/1200 & 31.578947 & 7.76978557(3)  \times 10^{-3}  &  3.2  \times 10^{-11}  &8.761556(3)  \times 10^{-3}  &  2.7  \times 10^{-9}  \\ 
37/1200 & 32.432432 & 7.31868525(3)  \times 10^{-3}  &  3.1  \times 10^{-11}  &8.289669(1)  \times 10^{-3}  &  1.2  \times 10^{-9}  \\ 
36/1200 & 33.333333 & 6.88365739(3)  \times 10^{-3}  &  2.9  \times 10^{-11}  &7.831639(1)  \times 10^{-3}  &  1.1  \times 10^{-9}  \\ 
35/1200 & 34.285714 & 6.46445084(3)  \times 10^{-3}  &  2.8  \times 10^{-11}  &7.387579(1)  \times 10^{-3}  &  1.2  \times 10^{-9}  \\ 
34/1200 & 35.294118 & 6.06081740(3)  \times 10^{-3}  &  2.7  \times 10^{-11}  &6.957224(1)  \times 10^{-3}  &  1.1  \times 10^{-9}  \\ 
33/1200 & 36.363636 & 5.67251178(3)  \times 10^{-3}  &  2.5  \times 10^{-11}  &6.540566(1)  \times 10^{-3}  &  1.0  \times 10^{-9}  \\ 
32/1200 & 37.500000 & 5.29929155(2)  \times 10^{-3}  &  2.4  \times 10^{-11}  &6.137787(1)  \times 10^{-3}  &  10.0  \times 10^{-10}  \\ 
31/1200 & 38.709677 & 4.94091706(2)  \times 10^{-3}  &  2.3  \times 10^{-11}  &5.748030(1)  \times 10^{-3}  &  1.0  \times 10^{-9}  \\ 
30/1200 & 40.000000 & 4.59715141(2)  \times 10^{-3}  &  2.1  \times 10^{-11}  &5.3720061(8)  \times 10^{-3}  &  8.1  \times 10^{-10}  \\ 
29/1200 & 41.379310 & 4.26776039(2)  \times 10^{-3}  &  2.0  \times 10^{-11}  &5.0093817(8)  \times 10^{-3}  &  7.7  \times 10^{-10}  \\ 
28/1200 & 42.857143 & 3.95251242(2)  \times 10^{-3}  &  1.9  \times 10^{-11}  &4.6600943(8)  \times 10^{-3}  &  7.5  \times 10^{-10}  \\ 
27/1200 & 44.444444 & 3.65117851(2)  \times 10^{-3}  &  1.8  \times 10^{-11}  &4.3240426(7)  \times 10^{-3}  &  7.5  \times 10^{-10}  \\ 
26/1200 & 46.153846 & 3.36353220(2)  \times 10^{-3}  &  1.7  \times 10^{-11}  &4.0011850(9)  \times 10^{-3}  &  8.8  \times 10^{-10}  \\ 
25/1200 & 48.000000 & 3.08934952(2)  \times 10^{-3}  &  1.6  \times 10^{-11}  &3.6914382(8)  \times 10^{-3}  &  8.4  \times 10^{-10}  \\ 
24/1200 & 50.000000 & 2.82840891(1)  \times 10^{-3}  &  1.5  \times 10^{-11}  &3.3947478(8)  \times 10^{-3}  &  8.3  \times 10^{-10}  \\ 
23/1200 & 52.173913 & 2.58049121(1)  \times 10^{-3}  &  1.4  \times 10^{-11}  &3.1110028(8)  \times 10^{-3}  &  8.1  \times 10^{-10}  \\ 
22/1200 & 54.545455 & 2.34537958(1)  \times 10^{-3}  &  1.3  \times 10^{-11}  &2.8401729(6)  \times 10^{-3}  &  6.5  \times 10^{-10}  \\ 
21/1200 & 57.142857 & 2.12285947(1)  \times 10^{-3}  &  1.2  \times 10^{-11}  &2.5821716(6)  \times 10^{-3}  &  6.2  \times 10^{-10}  \\ 
20/1200 & 60.000000 & 1.91271853(1)  \times 10^{-3}  &  1.1  \times 10^{-11}  &2.3369429(6)  \times 10^{-3}  &  6.0  \times 10^{-10}  \\ 
19/1200 & 63.157895 & 1.71474662(1)  \times 10^{-3}  &  9.7  \times 10^{-12}  &2.1043817(6)  \times 10^{-3}  &  5.8  \times 10^{-10}  \\ 
18/1200 & 66.666667 & 1.528735710(9)  \times 10^{-3}  &  8.9  \times 10^{-12}  &1.8844516(6)  \times 10^{-3}  &  5.6  \times 10^{-10}  \\ 
17/1200 & 70.588235 & 1.354479837(8)  \times 10^{-3}  &  8.0  \times 10^{-12}  &1.6770700(5)  \times 10^{-3}  &  5.4  \times 10^{-10}  \\ 
16/1200 & 75.000000 & 1.191775072(7)  \times 10^{-3}  &  7.2  \times 10^{-12}  &1.4821730(5)  \times 10^{-3}  &  5.2  \times 10^{-10}  \\ 
15/1200 & 80.000000 & 1.040419450(6)  \times 10^{-3}  &  6.5  \times 10^{-12}  &1.2996682(5)  \times 10^{-3}  &  5.0  \times 10^{-10}  \\ 
14/1200 & 85.714286 & 9.00212925(6)  \times 10^{-4}  &  5.8  \times 10^{-12}  &1.1295048(5)  \times 10^{-3}  &  4.7  \times 10^{-10}  \\ 
13/1200 & 92.307692 & 7.70957309(5)  \times 10^{-4}  &  5.1  \times 10^{-12}  &9.716037(4)  \times 10^{-4}  &  4.5  \times 10^{-10}  \\ 
12/1200 & 100.000000 & 6.52456222(4)  \times 10^{-4}  &  4.4  \times 10^{-12}  &8.258872(4)  \times 10^{-4}  &  4.2  \times 10^{-10}  \\ 
11/1200 & 109.090909 & 5.44515031(4)  \times 10^{-4}  &  3.8  \times 10^{-12}  &6.922942(4)  \times 10^{-4}  &  4.0  \times 10^{-10}  \\ 
10/1200 & 120.000000 & 4.46940790(3)  \times 10^{-4}  &  3.2  \times 10^{-12}  &5.707373(4)  \times 10^{-4}  &  3.7  \times 10^{-10}  \\ 
9/1200 & 133.333333 & 3.59542179(3)  \times 10^{-4}  &  2.7  \times 10^{-12}  &4.611449(4)  \times 10^{-4}  &  3.5  \times 10^{-10}  \\ 
8/1200 & 150.000000 & 2.82129437(2)  \times 10^{-4}  &  2.2  \times 10^{-12}  &3.634404(3)  \times 10^{-4}  &  3.2  \times 10^{-10}  \\ 
7/1200 & 171.428571 & 2.14514294(2)  \times 10^{-4}  &  1.7  \times 10^{-12}  &2.775443(3)  \times 10^{-4}  &  2.9  \times 10^{-10}  \\ 
6/1200 & 200.000000 & 1.56509892(1)  \times 10^{-4}  &  1.3  \times 10^{-12}  &2.033781(3)  \times 10^{-4}  &  2.6  \times 10^{-10}  \\ 
5/1200 & 240.000000 & 1.07930709(1)  \times 10^{-4}  &  9.8  \times 10^{-13}  &1.408601(2)  \times 10^{-4}  &  2.3  \times 10^{-10}  \\ 
4/1200 & 300.000000 & 6.85924588(7)  \times 10^{-5}  &  6.6  \times 10^{-13}  &8.99121(2)  \times 10^{-5}  &  2.1  \times 10^{-10}  \\ 
3/1200 & 400.000000 & 3.83119949(4)  \times 10^{-5}  &  4.0  \times 10^{-13}  &5.04335(2)  \times 10^{-5}  &  1.8  \times 10^{-10}  \\ 
2/1200 & 600.000000 & 1.69071800(2)  \times 10^{-5}  &  2.0  \times 10^{-13}  &2.23520(5)  \times 10^{-5}  &  4.6  \times 10^{-10}  \\ 
1/1200 & 1200.000000 & 4.19673156(6)  \times 10^{-6}  &  6.1  \times 10^{-14}  &5.571(6)  \times 10^{-6}  &  5.8  \times 10^{-9}  \\ 
\end{longtable*}
\end{widetext}
\raggedright
\bibliography{journalshortnames,references}

\end{document}